\documentclass[aps,twocolumn,nofootinbib,showpacs,floatfix]{revtex4}

\usepackage[centertags]{amsmath}
\usepackage{amsfonts}
\usepackage{graphicx}
\usepackage[hypertex]{hyperref}
\usepackage{psfrag}

\newcommand{\bea}{\begin{eqnarray}}
\newcommand{\eea}{\end{eqnarray}}
\newcommand{\be}{\begin{equation}}
\newcommand{\ee}{\end{equation}}
\newcommand{\av}[1]{\ensuremath{\left\langle#1\right\rangle}}

\newcommand{\ve}[1]{\ensuremath{\boldsymbol{#1}}}
\newcommand{\D}[1][ ]{\ensuremath{\mathrm{d}^{#1} }}
\newcommand{\tdyn}{\ensuremath{\tau_\text{dyn}}}

\begin{document}

\title{
Gravitational Dynamics of an Infinite Shuffled Lattice: 
Particle Coarse-grainings, Non-linear Clustering and the Continuum
Limit}

\author{T. Baertschiger}
\affiliation{Dipartimento di Fisica, Universit\`a ``La Sapienza'',
P.le A. Moro 2,
I-00185 Rome,
Italy,\\
\& ISC-CNR,
Via dei Taurini 19,
I-00185 Rome,
Italy.}
\author{M. Joyce}
\affiliation{Laboratoire de Physique Nucl\'eaire et de Hautes Energies,
Universit\'e Pierre et Marie Curie-Paris 6,
UMR 7585,
Paris, F-75005 France.}
\author{A. Gabrielli}
\affiliation{ISC-CNR,
Via dei Taurini 19,
I-00185 Rome,
Italy,\\\& SMC INFM/CNR,
Dipartimento di Fisica,
Universit\`a ``La Sapienza'',
P.le A. Moro 2,
I-00185 Rome,
Italy.}
\author{F. Sylos Labini}
\affiliation{ ``E. Fermi'' Center, Via Panisperna 89 A, Compendio del
Viminale, I-00184 Rome, Italy,\\
\& ISC-CNR,
Via dei Taurini 19,
I-00185 Rome,
Italy.}
\begin{abstract}    
\begin{center}    
{\large\bf Abstract}    
\end{center}     
We study the evolution under their self-gravity of particles
evolving from infinite ``shuffled lattices'' initial
conditions. We focus here specifically on the comparison
between the evolution of such a system  
and that of ``daughter'' coarse-grained particle distributions.  These are
sparser (i.e. lower density) particle distributions, defined by a
simple coarse-graining procedure, which share the same large-scale 
mass fluctuations. We consider both the case that such
coarse-grainings are performed (i) on the initial conditions, and (ii)
at a finite time with a specific additional prescription.  In
numerical simulations we observe that, to a first approximation, these
coarse-grainings represent well the evolution of the two-point
correlation properties over a significant range of scales. We note, in
particular, that the form of the two-point correlation function in the
original system, when it is evolving in the asymptotic
``self-similar'' regime, may be reproduced well in a daughter
coarse-grained system in which the dynamics are still dominated by
two-body (nearest neighbor) interactions. This provides a simple
physical description of the origin of the form of part of the
asymptotic non-linear correlation function.  Using analytical results
on the early time evolution of these systems, however, we show that
small observed differences between the evolved system and its
coarse-grainings {\it at the initial time} will in fact diverge as the
ratio of the coarse-graining scale to the original inter-particle
distance increases. The second coarse-graining studied, performed at a
finite time in a specified manner, circumvents this problem.  It also
makes more physically transparent why gravitational dynamics from
these initial conditions tends toward a ``self-similar'' evolution.
We finally discuss the precise definition of a limit in which a 
continuum (specifically Vlasov-like) description of the observed 
linear and non-linear evolution should be applicable. This requires
the introduction of an additional intrinsic length scale (e.g.
a physical smoothing in the force at small scales), which is
kept fixed as the particle density diverges. In this limit the
different coarse-grainings are equivalent and leave the 
evolution of the ``mother'' system invariant.
\end{abstract}    
\maketitle    
\date{today}    

\twocolumngrid


\section{Introduction}
\label{Introduction}

The dynamics of infinite self-gravitating systems of point particles starting
from quasi-uniform initial conditions is a straightforward, well posed, but
essentially unsolved problem of out of equilibrium statistical mechanics.
While numerical simulation of such systems has developed impressively
in scale and sophistication in the last few decades  --- mostly driven by 
the relevance of the problem in cosmology ---  analytic understanding of
the evolution of the clustering observed remains very limited. The work
reported in this article is a continuation of a study of this problem
reported by us recently \cite{sl1}. We study a simplified version of the 
problem posed in cosmology,  considering pure Newtonian gravity in a 
static euclidean universe and a very simple class of ``shuffled lattice'' 
initial conditions (IC). 

In \cite{sl1} we have presented in detail the results of a numerical
study of this system. We have found that, as soon as strong positive
two-point correlations begin to develop, they are characterized by a
simple spatio-temporal scaling relation, in which the strong
spatial correlations at any time may be inferred from those at a
precedent time by a simple rescaling of the spatial variables. Further
the function of time specifying this dynamical scaling of the
characteristic spatial scale tends, after a transient period, to a
form which can be determined solely from the linearized fluid
equations describing the evolution of the small fluctuations at large
scales. This asymptotic dynamical scaling behavior is completely
analogous to that which has been observed in cosmological simulations,
starting from more complicated IC (correlated perturbations of
lattices) and in an expanding universe, and referred to in this latter
context as ``self-similarity''.  The amplitude of the two-point
correlation function is also observed in this asymptotic regime to be
essentially independent of the mean particle density over a large
range of scale. This suggests that the full non-linear dynamics of
these simulations, once this asymptotic scaling behavior is
approached, might be well approximated solely within the framework of
a continuous fluid-like description of the system.  We have found,
however, that the {\it form} of the two-point correlation function
(i.e. its dependence on separation rather than its temporal scaling
behavior) attained in this asymptotic scaling regime is already very
well approximated, at least for modest amplitudes, by that at early
times where the dynamics at these scales is explicitly particle-like
in nature. Indeed the correlation function which develops at these
times is very well approximated by assuming that the dynamics is that
of particles interacting only with their few nearest neighbor
particles.  In other words, while the spatio-temporal scaling law
seems to be determined by only the large scale fluid linear dynamics,
the form of the scaling function looks to depend on the discrete
aggregation dynamics between elementary objects.  
This suggests that the discreteness of the system may be an essential
ingredient even to understand its long-time behavior.

In this paper we study further the dynamical evolution in this same class of
models (i.e. with ``shuffled lattice'' IC in a static universe). 
Specifically we focus here on the comparison between the evolution of a
given such IC and the evolution of other point distributions defined
from them by a simple coarse-graining procedure: at a chosen time 
the mass in coarse-graining cells, defined by a lattice, is aggregated 
into a single mass; this new point mass is attributed the position and 
velocity of the center of mass of the mass in the cell.  This is
a procedure which evidently  modifies the system up to a certain 
scale --- that characteristic of the coarse-graining --- but leaves 
the mass and velocity fluctuations at larger scales essentially
unperturbed. The comparison
of the original and coarse-grained (CG) system allows one 
to infer useful physical information about the dynamics. Specifically
it allows one to study the role of fluctuations at different scales
in the system, and the degree to which a discrete or continuous description of
the dynamics can be appropriate. We will see that the study will allow
us to understand better the physics underlying the asymptotic
``self-similar'' evolution, and also to address more precisely the 
question of the continuum limit of the system.  These latter 
considerations are of interest, as we will explain in the conclusions, 
in the context of simulations in cosmology. Such simulations can be seen 
as particle coarse-grainings in the sense defined here of the ``true'' 
self-gravitating matter described by cosmological 
models\footnote{For typical ``cold dark matter'' models  there 
is a factor of the order of $10^{70}$ between the microscopic 
number density of the physical particles and the number density 
of the ``macro-particles'' used in the largest simulations performed
up to now.}. A clear 
understanding of the continuum limit is important as it should 
correspond to the theoretical cosmological model. Discreteness effects 
in these simulations are the differences between the CG and this
underlying continuum model. 

In the introduction to \cite{sl1} we have given a brief review of the
relevant literature, on both self-gravitating systems and other
long-range interacting systems. The notion of ``particle
coarse-graining'', which is the central one in this paper, is simply a
specific kind of coarse graining procedure introduced in
statistical physics (see, e.g.\cite{cardy}). It is evidently not unnatural to
consider such schemes in approaching the dynamics of systems
manifesting scaling behavior, as has been done widely in many
different systems in condensed matter physics with well known
results. In the context of self-gravitating systems various authors
have studied self-gravitating dynamics applying such concepts (see,
e.g., \cite{gaite_dominguez_scaling, gaite_proc2000_renormalization,
Sota_etal_renormalization, Semelin_renormalization, Antonov_scaling}). 
These works, however, study theoretical descriptions of these systems in a
continuum limit. In this work we consider, instead, the fully discrete
systems, numerically simulated, applying coarse-graining procedures
which relate one such discrete system to another. Indeed one of the
questions we address in this work is, as mentioned above, the validity
of a continuum description of the dynamics. One of our findings is
that the coarse-grained system, in a phase in which its dynamics are
manifestly discrete, already traces well the behavior of the original
system.  The only previous work we are aware of which has actually
applied a discrete ``renormalization'' procedure is that of
\cite{Peebles_renormalization, Peebles+couchman_1995}, which attempts
to determine the mass distribution which results from power law
initial conditions using such a scheme under the {\it assumption} that
the evolution is self-similar. Numerical studies of discreteness
effects in cosmological N body simulations (e.g. \cite{splinter}) also
implicitly consider the effect of different discrete coarse-grainings,
albeit not defined as such in the simple way considered here. Indeed,
as mentioned above, one of the goals of this work is to clarify
problems in this context by a well controlled study of much simpler
systems.

The paper is organized as follows. In the next section we summarize
briefly the essential definitions and the results of the preceding
paper \cite{sl1}. We also summarize relevant results derived recently
by the present authors and B. Marcos \cite{marcos_06,joyce_05} on a
perturbative description of the early times dynamics of
self-gravitating points perturbed off a lattice. We underline the
slight differences in the systems studied here compared to that in
\cite{sl1} --- a different probability distribution function for the
initial ``shuffling'' and also a different ascription of initial
velocities. To explain this latter point we include a short discussion
of the ``Zeldovich approximation'' to the evolution of a pressure-less
self-gravitating fluid (This approximation is also important for the
discussion in Sec.~\ref{sec:later_time_evol} of the paper). In
Sec.~\ref{sec:CG_procedure} we define precisely and discuss the
construction of the coarse-grained particle distributions from a given
initial particle system. In the following section we report our study
of the comparison between a given SL and a set of CG of the initial
conditions, while in Sec.~\ref{sec:later_time_evol} we do the same
for a set of CG defined at a given scale at finite time in the course
of the evolution.  In the final section we summarize our results and
conclusions, and discuss what we have learned about the dynamics of
these systems. Specifically we discuss the question of the appropriateness
of a continuum description of the observed non-linear dynamics, and
we indicate some paths for future research on this question.

\section{Basic definitions and background results}
\label{Basic definitions and background results}

In this section we give the essential definitions and background for
the paper. Firstly we summarize the basic definitions of the initial
conditions studied and of the quantities used to characterize the
systems. Except for a few minor points this discussion essentially
summarizes an analogous one in \cite{sl1}, where the interested reader
can find further details. One minor difference is that we consider
here the analysis also of particle distributions in which the masses of
all points are not equal. We then summarize the essential results of
\cite{sl1}, and then also those of \cite{marcos_06} which we also make
use of in this paper. Finally, we discuss briefly the so-called
``Zeldovich approximation'' which is an important perturbative
approximation to the evolution of self-gravitating systems in the
fluid limit, and therefore an approximation to the evolution of the
particle system in a certain regime (long wavelengths and small
amplitude perturbations). We will use it specifically, as it is
commonly done in cosmological simulations, in the choice of initial
conditions for the particle velocities, and also at various points in
our analysis.

\subsection{Definition of a Shuffled Lattice} 
\label{Definition of a Shuffled Lattice} 

We call \emph{shuffled lattice} (SL) the particle distribution
obtained by applying a random displacement independently to 
each particle on a simple cubic lattice. Thus the
position of a particle, initially on the lattice site $\ve R$,
is given by $\ve x(\ve R) = \ve R + \ve u(\ve R)$ where the 
vectors $\ve u(\ve R)$ are random and specified by
a probability density functional 
\begin{equation}
\mathcal{P}[\left\{ \ve u(\ve R)\right\}] = \prod_{\ve R} p[\ve u(\ve
R)]  \ .
\end{equation}
The statistical properties of the particle distribution are thus completely 
determined by $p(\ve u)$, the probability density function (PDF)
for the displacement of a single particle.

In the numerical simulations reported in \cite{sl1} we took
a simple ``box'' form for the PDF. Here instead, for reasons
which we will explain below,  we will take $p(\ve u)$ 
to be simply a mono-variate Gaussian:
\begin{equation}
p(\ve u) \equiv p(u) = \left( \frac{1}{2\pi \sigma^2}\right)^{3/2} 
 \exp\left( - \frac{u^2}{2\sigma^2}\right)  \,,
\label{eq:pu}
\end{equation}
where $u$ is the modulus of $\ve u$. As discussed in \cite{sl1} we 
expect the choice of the precise
form of the PDF (for any simple functional form) to have little 
effect on the dynamics beyond some early time transients. 

We use the term \emph{shuffling parameter}, denoted $\Delta$, for 
the rms deviation of $\ve u$, i.e.,
\begin{equation}
\Delta^2 = \int_{\mathbb R^3} u^2\  p({\ve u})\, \mathrm d^3u =
3 \sigma^2 \ . 
\label{eq:defDelta}
\end{equation} 
We will usually use this quantity expressed in terms of the lattice
spacing, which we denote by $\ell$. We thus define the
\emph{normalized shuffling parameter} $\delta \equiv \Delta / \ell$.  
This is convenient because 
it is actually the value of $\delta$ alone which characterizes 
the gravitational dynamics of an infinite SL \cite{sl1}: any 
two SL with the same $\delta$ but different $\ell$ 
(and $\Delta$) are equivalent up to a scale transformation 
which is irrelevant for (scale-free) gravity. 
We note that in the limit $\delta\to 0\,$ the 
particle distribution remains a perfect cubic lattice, while 
for $\delta\to \infty$ it becomes a Poisson distribution. 

\subsection{Characterization of particle distributions}
\label{characterisation}

We now discuss briefly the different statistical quantities which we
will use to characterize the SL initial conditions and the evolved
particle distributions. For further details, we refer the reader to
\cite{sl1, gabrielli_06, book}.

For any distribution of $N$ particles in a volume $V$, 
we can define the microscopic mass density as
\begin{equation}
\rho (\ve x) = \sum_{i=1}^N m_i \delta_\text{D} ( \ve x - \ve x_i)
\label{eq:micro}
\end{equation}
where $\ve x_i\in V$ is the position of the $i$th particle, of mass $m_i$,
and $\delta_\text{D}$ is the Dirac delta function. For the case
of infinite systems with a well defined non-zero mean density
 $\rho_0$, which is that we consider here, it is convenient to 
define the density contrast $\delta_\rho (\ve x) = [\rho(\ve x) -
\rho_0]/\rho_0$.

The two-point correlation properties can then be characterized
by the reduced two-point (density-density) correlation function
\begin{equation}
\tilde{\xi}_\rho (\ve r) = \av{ \delta_\rho(\ve r + \ve x) \delta_\rho(\ve x) }\end{equation}
where $\av{\ldots}$ is an ensemble average. 
In our simulations we treat $N$ particles in a box of side
$L=N^{1/3} \ell$ with periodic boundary conditions.
We can therefore write the density contrast as a Fourier series:
\begin{equation}
\delta_\rho (\ve x) = \frac{1}{L^3} \sum_{\ve k} \exp(i\ve k\cdot \ve x)
\tilde{\delta}_\rho(\ve k)
\label{eq:deltax}
\end{equation}
with $\ve k \in \{ (2\pi/L)\ve n | \ve n\in \mathbb Z^3\}$, and 
\begin{equation}
\begin{split}
\tilde{\delta}_\rho (\ve k) &= 
\int_{L^3} \exp(-i\ve k\cdot \ve x) \delta_\rho (\ve x) \\ 
&=\frac{1}{\rho_0} \left[ \sum_i m_i \exp(-i \ve k\cdot \ve
 x_i) -  \delta_\text{K}(L\ve k/2\pi,\ve 0)\sum_i m_i \right] \ , 
\label{eq:deltak}
\end{split}
\end{equation}
where $\delta_\text{K}$ is the Kronecker delta function.
The power spectrum\footnote{As in \cite{sl1} we use the terminology
and definitions current in cosmology. In solid state physics the PS is
usually called the structure factor, and normalized typically 
so that it is dimensionless rather than having units of volume.}
(PS) may then be defined (see, e.g. \cite{book}) as 
\begin{equation}
P(\ve k) = \frac{1}{L^3} \av{|\tilde\delta_\rho (\ve k)|^2} \ . 
\end{equation}
For statistically homogeneous (i.e. statistically translational invariant)
particle distributions\,\footnote{In a lattice, the
  ensemble average is defined over the set of lattices translated by
  an arbitrary vector in the lattice cell. In the case of the SL
we have both this average and, in addition, that over all realizations
 of $\mathcal P[\{ \ve
  u(\ve R)\}]$.}, the PS and $\tilde \xi(\ve r)$ are a Fourier
conjugate pair and therefore contains the same information.  

We estimate the PS as a function of the modulus $k=|\ve k|$,
using simply the expression given in Eq.~(\ref{eq:deltak}), 
averaged over all $\ve k$ in a shell in $k$ space. We
therefore denote it from now on as $P(k)$. For the real
space correlations we will consider estimating 
$\tilde{\xi}_\rho (\ve r)$ as defined above only in the 
cases where the points of the distribution considered
have equal mass. Just as for the PS we will evaluate it
by performing an average over all $\ve r$ in a radial
shell. In this case we will denote it as $\xi (r)$, and
it can be written\cite{book}, 
for $\ve r \neq 0$, as 
\begin{equation}
\xi(r) = \frac{\av{n(r)}_p}{n_0} -1  
\label{xi-r}
\end{equation}
where $\av{n(r)}_p$ is the radial conditional mean number
density (i.e. the mean number density of particles at a distance $r$
from an occupied point), and $n_0$ is the mean (unconditioned) number
density.  This makes its direct estimation in real space very simple,
by simple pair-counting algorithms (see \cite{sl1} for the explicit
description of the algorithm we choose). 

Besides these two quantities we will consider also, as in
\cite{sl1}, one other one: the nearest neighbor
(NN) distribution $\omega(r)$, which is simply the probability
density function for a particle to have its NN at a distance $r$.
As in \cite{sl1} it is estimated in the evident way by
direct counting of NN separations. The usefulness of this 
quantity here is that it allows one to infer \cite{Baertschiger:2004tx}
very useful information about the nature of the correlations
at small scales,
and thus indirectly about the dynamics responsible for them:
if we neglect all but the two-point correlations one has 
the relation \cite{book}
\begin{equation} 
\label{omega1} 
\omega(r)  = \left( 1 - \int_0^r\omega(s)\, \D s \right) \cdot
4 \pi r^2 n_0  [ 1+ \xi(r) ]  \;.
\end{equation}
We will use this relation (as in \cite{Baertschiger:2004tx, sl1})
to probe the degree to which interactions with NN particles are
responsible at early times in simulations for the development
of the observed correlations.

	
\subsection{Gravitational clustering in a Shuffled Lattice}
\label{Gravitational clustering in a Shuffled Lattice}

It is straightforward \cite{andrea,book,gabrielli_06} to calculate 
exactly the  two-point correlation properties of the SL. The
PS is given by 
\begin{equation}
 P(\ve k) = \frac{1-|\tilde{p}(\ve k)|^2}{n_0} + L^3 \sum_{\ve n}
 |\tilde{p}(\ve k)|^2 \,  \delta_\text{K}( \ve k , \ve n \frac{2 \pi}{\ell}) 
\label{eq:exactPS}
\end{equation}
where $\tilde p(\ve k)$ is the Fourier transform of the 
PDF for the displacements $p(\ve u)$ (i.e. its characteristic 
function). For the case of a Gaussian PDF as in Eq.~\eqref{eq:pu},
with Eq.~\eqref{eq:defDelta}, it follows that, for 
$\ve k \neq {\ve n} \frac{2\pi}{\ell}$,
\begin{equation}
P(\ve k) \equiv P(k)=\frac{1}{n_0}
\left[1-\exp \left(-\frac{\Delta^2 k^2}{3}\right)\right]\,.
\label{psSL-gaussian}
\end{equation}
At leading order in small $k$ we therefore have 
\begin{equation}
P(\ve k) \equiv P(k)=\frac{\Delta^2 k^2}{3n_0}\,.
\label{psSL-leading term}
\end{equation}
Note that, at large $\ve k$, one has $P(\ve k) = 1/n_0$,
which is simply the particle shot noise necessarily present
in any distribution of point-particles.

In \cite{sl1} we have studied the dynamical evolution of 
self-gravitating particles starting from SL initial
conditions, and zero initial particle velocities. The 
PDF $p(\ve u)$ used for the initial displacements off the lattice
is a simple ``box'' form, i.e., with uniform probability inside
a cube centered on the origin  and oriented parallel
to the lattice.  The leading term in the 
PS Eq.~(\ref{psSL-leading term}) is in fact independent of 
the form\footnote{The only assumption about
the PDF is that is has finite variance. The qualitatively
different case of infinite variance PDF is treated also
in \cite{andrea, book}.} of $p(\ve u)$, and thus in this approximation the 
initial conditions are the same.  

The principal results of \cite{sl1} are the following:

\begin{itemize}
\item Qualitatively the evolution of clustering is very similar
to that observed in cosmological simulations 
(see, e.g., \cite{efstathiou_88}), which are 
performed in an expanding spatial background and start from 
lattices with {\it correlated} displacements representing PS 
typically with  $P(k) \propto k^n$ and $-3<n<-1$. Clustered 
structures develop initially at very small scales ($< \ell$) 
and then progressively at larger and larger scales.
 
\item The evolution is observed to be independent of the box size
$L$ until the time when the size of the non-linear structures 
approaches the box size itself. Thus the results of numerical simulations
are interpreted as representative of the infinite volume limit
(taken at constant particle density).

\item  The PS at small $k$ is amplified as predicted by the linearized
fluid limit for the system (see, e.g. \cite{peebles}). The $k$ 
below which this behavior is observed decreases with time,
reflecting the propagation of the non-linear clustering in real space
to larger scales.

\item The temporal evolution of two-point correlation function 
which coincides in this case (of equal mass points) with the 
density-density correlation function, is well described from 
early times by a {\it spatio-temporal scaling relation} 
\be
\label{scaling_general}
\xi(r,t)=f \left(\frac{r}{\lambda(t)} \right)\,,
\ee
i.e., the temporal evolution of $\xi(r,t)$ is well approximated 
by a simple rescaling of the spatial coordinates\footnote{In the 
rest of the paper we will often write $\xi(r)$ leaving the 
time dependence implicit.}.

\item After a time {\bf $\Delta t$} this scaling behavior converges toward 
a more specific form which may be written
\begin{equation} 
\xi (r, t + \Delta t) = \xi \left(\frac{r}{R_s(\Delta t)}, t \right)
\; ; \quad R_s(\Delta t) = e^{ z \frac{\Delta t}{\tdyn} } \,,
\label{eq:self-sim-xi}
\end{equation}
where $\tdyn$ is the {\it dynamical time} of the system, defined by
$\tdyn=1/\sqrt{4\pi G \rho_0}$. The constant $z$, which one can consider
a sort of {\it dynamical exponent} characterizing the dynamical scaling,
is given by
\be
z=\frac{2}{3+n} 
\ee
where $n$ is the exponent of the power-law PS of the initial particle
distribution (and therefore $z=2/5$ for the SL).  This behavior is
completely analogous to what is referred to as ``self-similarity'' in
the context of cosmology (see, e.g., \cite{efstathiou_88}). It is 
the behavior which follows if one
assumes that (i) the characteristic length scales in the system play
no role in the dynamics, and (ii) the linearized fluid limit correctly
describes {\bf \it the evolution of} fluctuations at low amplitude.
It is thus expected to be observed specifically starting from IC which
are power law PS, if the clustering is independent of the lattice
spacing (which is, in this case, the sole characteristic length
scale). The results in \cite{sl1} extend the range in which such
behavior has been reported in numerical simulations in cosmology to
``bluer'' PS (and also to the case of a static universe).

\item We have mentioned that the sole relevant parameter for
the gravitational dynamics of a SL is the normalized shuffling
parameter $\delta$. In \cite{sl1} different SL with a range
of $\delta$, but identical large scale amplitude of the PS,
are compared. The evolution of these IC are all observed to
converge to the self-similar behavior after a transient 
which depends on $\delta$. These results are closely related to 
those described in Sec.~\ref{CG at t=0} below. 

\item In the early phase of evolution, in which non-linear 
clustering develops, but prior to the onset of the self-similarity,
the observed two-point correlations are very well accounted for
solely by two body correlations developing under the influence of 
NN interactions. The form of the two-point correlation
function at this time is also a rather good approximation 
to that in the self-similar regime. Indeed, as noted
above, the spatio-temporal scaling of the correlation function
is a good approximation well before the self-similar regime 
is attained, i.e., in the transient phase we have  
already a temporal evolution given approximated well by
Eq.~(\ref{scaling_general}), but with a different temporal 
dependence of the function $\lambda(t)$ to that in the 
asymptotic regime of dynamical scaling 
Eq.~(\ref{eq:self-sim-xi}).

\end{itemize}

The aim of this paper, as has been discussed in the introduction,
is to gain more insight into the dynamics at work in these
simulations by analyzing the effect of coarse-graining the
system at different times in the course of its evolution.  

\subsection{Early time evolution of a perturbed lattice}
\label{PLT}

In \cite{joyce_05, marcos_06} we have developed in detail a general 
formalism for treating the evolution of self-gravitating particles 
perturbed off an 
infinite perfect lattice, in the limit that these 
perturbations are such that the relative displacements of the points 
are small compared to their separations. The treatment is thus valid
up to times at which this approximation breaks down, and is 
applicable to any set of perturbations, whether uncorrelated (as
considered here) or correlated (as in cosmological simulations).
It applies equally well to the case of a static euclidean universe
or a cosmological expanding one. We will make use of these results
below and thus summarize them briefly here. We refer to this 
approximation to the full dynamics as {\em particle linear theory}
(PLT). 

The formalism is essentially the same as that used canonically
to analyze classical phonons in a crystal in solid state physics
(see e.g. \cite{pines}). One can write the equation of motion for 
the displacement $\ve u(\ve R)$  of a particle, from its 
lattice site $\ve R$, as:
\be
{\bf {\ddot u}}({\bf R},t) = -\sum_{{\bf R}'} 
{\cal D} ({\bf R}- {\bf R}') {\bf u}({\bf R}',t)\,. 
\label{linearised-eom}
\ee
where  ${\cal D}$ is the {\it dynamical matrix}, which is derived
by linearizing the force on a particle in the displacements 
relative to all other particles. For gravity 
we have 
\begin{eqnarray}
{\cal D}_{\mu \nu} ({\bf R} \neq {\bf 0}) &=&
Gm\left[\frac{\delta_K (\mu, \nu)}{R^3}
-3\frac{R_\mu R_\nu}{R^5}\right]\\
{\cal D}_{\mu \nu} ({\bf 0}) &=& 
-\sum_{{\bf R} \neq {\bf 0}} {\cal D}_{\mu \nu} ({\bf R})\,.
\label{Dmatrix-gravity}
\end{eqnarray}
Defining the discrete Fourier transform on the lattice by 
\begin{eqnarray}
{\bf {\tilde u}}({\bf k},t)= \sum_{{\bf R}} e^{-i {\bf k}\cdot{\bf R}}
{\bf u}({\bf R},t) \\
{\bf u}({\bf R},t)= \frac{1}{N} \sum_{{\bf k}} e^{i {\bf k}\cdot{\bf R}}
 {\bf {\tilde u}}({\bf k},t)
\end{eqnarray}
where the sum in $\ve k$ is over the first Brillouin zone, i.e., for
the simple cubic lattice ${\ve k} \in {\ve n} (2\pi /L)$, where $\ve n$
is a vector of integers each  
$\in ]-\frac{N^{1/3}}{2} \;, \frac{N^{1/3}}{2}]$, the
equation of motion Eq.~(\ref{linearised-eom}) becomes, in reciprocal
space,
\be
{\bf \ddot{{\tilde u}}} ({\bf k},t) 
= -{\cal {\tilde D}} ({\bf k}) {{\bf {\tilde u}}}({\bf k},t) 
\label{linearized-eom-kspace}
\ee
where ${\cal {\tilde D}} ({\bf k})$, the FT of 
${\cal D} ({\bf R})$, is a real symmetric $3 \times 3$ matrix for 
each ${\bf k}$.  

The dynamical problem thus reduces to the diagonalisation of ${\cal
{\tilde D}} ({\bf k})$ for each ${\bf k}$ in the first Brillouin
zone. Labeling the three orthonormal eigenvectors ${\bf {\hat e}}_n ({\bf
k})$ and their eigenvalues $\omega_n^2({\bf k})$ ($n=1,2,3$), the
evolution of the displacement field from $t=0$ is given as\footnote{We
give here the formulae for the specific case of a static euclidean
universe. The general result for cosmological backgrounds is given
also in \cite{marcos_06}.}
\bea
{\bf u}({\bf R}, t) &=\frac{1}{N} 
\sum_{{\bf k}} \sum_{n}
 {\bf {\hat e}}_n({\bf k}) \cdot \{ {\bf {\tilde u}}({\bf k},0)\cdot{\bf {\hat e}}_n({\bf k}) 
\cosh [\omega_n ({\bf k}) t]
\nonumber \\
&+ {\bf \frac{1}{\omega_n ({\bf k})} \dot{\tilde u}} ({\bf k},0)\cdot{\bf {\hat e}}_n({\bf k})  
\sinh [\omega_n ({\bf k}) t] \} e^{i {\bf k}\cdot{\bf R}}
\label{linearised-evolution-general}
\eea
Thus given the initial displacements and velocities, the dynamical 
evolution is solved. 

The solution of the diagonalisation problem is numerically 
straightforward. Details of it, and the results for the eigenvalues and 
eigenvectors, are given in \cite{marcos_06}. In this paper 
the domain of validity of this perturbative treatment has also
been investigated  using numerical simulations. Starting from (i) 
an SL initial condition and (ii) from a perfect lattice configuration 
with correlated perturbations corresponding to a PS $\sim k^{-2}$, it 
is found that the treatment traces very well, in both cases, the 
full evolution of the system (i.e. with the full gravitational interaction) 
until the average relative displacement approaches the
lattice spacing. 

\subsection{The Zeldovich Approximation}
\label{The Zeldovich Approximation}

Keeping the mean mass density $\rho_0$ fixed, and taking the 
limit $\ell \rightarrow 0$, the eigenvectors and eigenvalues 
at any given fixed $k$ simplify: one obtains a single non-zero eigenvalue
{\it independent of k}, $\omega^2({\bf k})= 4\pi G \rho_0 \equiv \tdyn^{-2}$,
associated to a longitudinal mode 
${\bf {\hat e}}({\bf k})={\hat {\bf k}}$, and two transverse modes
(i.e. ${\hat e} ({\bf k}) \cdot {\hat {\bf k}}=0$) with zero
eigenvalues. If we consider the case that the perturbations to
the lattice are of long wavelength, i.e., that there are 
perturbations only with $k \ell \ll 1$, we obtain from
Eq.~(\ref{linearised-evolution-general}) that there is
an asymptotic attractor solution for large times ($t \gg \tdyn$),
which may be written
\be
{\bf u}({\bf R}, t) = \exp (t/\tdyn)\, {\bf q}({\bf R})\,,
\label{ZA}
\ee
where ${\bf q}({\bf R})$ is the time independent curl-free
(irrotational) vector field:
\be
{\bf q}({\bf R})= \frac{1}{N} \sum_{{\bf k}} 
\left\{
({\bf {\tilde u}}({\bf k},0) \cdot 
{\hat {\bf k}}) + 
\tdyn 
%
({\bf  \dot{\tilde u}}   
({\bf k},0) \cdot {\hat {\bf k}})
\right\} 
\cdot {\hat {\bf k}} e^{i {\bf k}\cdot{\bf R}}
\label{ZA-vector}
\ee
Using Eq.~(\ref{linearized-eom-kspace}) it is easy to show that the
gravitational field can be written, in the same long-wavelength limit,
as 
\be
{\bf g} ({\bf R}, t) = \tdyn^{-2} {\bf u} ({\bf R}, t) 
\ee
so that it follows that the asymptotic solution can be written
as   
\bea
{\bf u} ({\bf R}, t) =  \exp (t/\tdyn) \tdyn^{2} {\bf g} ({\bf R}, 0) \,, 
\label{ZA-grav-field-1} \\
\dot{\bf u} ({\bf R}, t) =  \exp (t/\tdyn) \tdyn {\bf g} ({\bf R}, 0) \,,
\label{ZA-grav-field-2}
\eea
i.e., both the displacement and velocity of each particle is expressed
solely in terms of the gravitational field acting on it at the initial
time \footnote{The approximation is thus sometimes described as
``ballistic''. This, however, is a misnomer as the approximation
incorporates in fact the collective effect of the motion of all
particles on one another: the approximation is non-trivial precisely
insofar as it extends well beyond the regime of a ballistic
approximation.}.

This approximation corresponds to one introduced by Zeldovich 
\cite{zeldovich_70}, for the evolution of fluid elements in a pressure-less 
self-gravitating fluid away from a perfectly uniform state, in
an expanding universe. It can
be formally derived \cite{buchert2} as an approximation through a 
perturbative treatment of the fluid equations in the Lagrangian 
formalism\footnote{The more general form of the solution at leading
order in this perturbative scheme can in fact be derived using the
formalism described here. See \cite{joyce_05,marcos_06} for details.}.

Note that using the continuity equation for small displacements
${\bf u}$ applied to elementary volumes in a continuous medium 
one obtains that the density fluctuations $\delta \rho$ with
respect to the mean density $\rho_0$ are given by 
$\delta \rho \approx -\rho_0{\bf \nabla }\cdot {\bf u}$. 
Linear amplification of small density fluctuations is thus
associated with the growth of displacements described by
Eq.~(\ref{ZA}), and indeed it was on the basis of this
observation that Zeldovich proposed his ansatz. The power of 
the Zeldovich Approximation (ZA) is that its domain of validity 
extends well beyond that 
of linear (Eulerian) perturbation theory of the self-gravitating
fluid. In fact it extends up to ``shell-crossing'', when fluid
elements contract in one direction to produce density
singularities. In terms of our particle treatment this 
shell-crossing corresponds to the approach of particles to one
another.

In cosmological simulations the ZA --- in the form of
Eq.~(\ref{ZA-grav-field-1}) adapted to an expanding universe ---
is used to set up initial conditions: it gives a prescription
for both the displacements and velocities of the particles
off the lattice (considered as the centers of fluid elements)
once the input PS is given, as any realization of the latter 
gives the gravitational field through the Poisson equation.
Our initial conditions are not equivalent to this as 
the displacement field applied to the lattice at the initial
time is not, in general, curl-free as in the ZA (since the 
gravitational field is the divergence of a scalar field).
We can, however, use the ZA to determine the velocities
since Eqs.~(\ref{ZA-grav-field-1}) and  (\ref{ZA-grav-field-2})
imply
\be
\dot{\bf u} ({\bf R}, 0) =  \frac{{\bf u} ({\bf R}, 0)}{\tdyn}\,.
\label{vel-IC}
\ee 
This is the prescription we will adopt here (rather than the zero
initial velocities of \cite{sl1}). It is a very natural one as it 
uses the single characteristic time scale of the fluid evolution 
to define the velocities. Further the choice is essentially 
equivalent to that in cosmological simulations, as it means 
that the PS at small $k$ (i.e. in the fluid limit) evolves 
exactly as in this context, growing in proportion to the 
square of the growing mode of linear theory $e^{t/\tdyn}$~\footnote{This
is true because in the small $k$ (fluid) limit only the curl-free 
component of the displacement field contributes to the density
fluid fluctuations. This follows from the relation
$\delta \rho \approx -{\bf \nabla }\cdot {\bf u}$, which is
valid in this limit. See \cite{joyce_05, marcos_06} for the exact 
expressions and analysis of the fluid limit.}, and likewise the 
displacement and velocity fields in the fluid limit:
\bea
{\bf u} ({\bf R}, t) &=& \exp (t/\tdyn) {\bf u} ({\bf R}, 0) 
\label{ZA-grav-field-3} \\
\dot{\bf u} ({\bf R}, t) &=& \exp (t/\tdyn) \dot{\bf u} ({\bf R}, 0)  \,.
\label{ZA-grav-field-4}
\eea


\section{Particle Coarse-grainings}
\label{sec:CG_procedure}

\subsubsection{Defining a coarse-graining (CG)}

In this section we describe precisely how we construct the coarse-grained
particle distributions we study. 

To coarse-grain a distribution of massive point particles:
\begin{itemize}

\item 1. We define a set of finite non-overlapping equal volume cells
covering all space, i.e., a tiling of three dimensional space with 
equal volume tiles. We consider here, for simplicity, only the 
case that these cells are those of a {\it simple cubic lattice}, with 
lattice spacing which we denote $\ell_{CG}$. Since the system studied
here is evolved assuming periodicity in a cubic box of side $L$, 
we choose $\ell_{CG}=L/p$ where $p$ is a positive integer. 

\item 2. We ascribe one particle to each cell. The mass of this (point)
particle is equal to the sum of the masses of the particles which are inside
the given cell in the distribution being coarse-grained, and its 
position and velocity coincide with those of the center of mass 
and of these particles.

\end{itemize}

We study here an initial particle
distribution which is a SL or a SL evolved under gravity, in which all
the points are ascribed equal mass. In general the 
point particles in the coarse-grained distribution just defined will
not have the same mass since the number of points in the
coarse-graining cells will not necessarily be equal. However in the 
cases which we consider in this paper, these particle mass fluctuations 
introduced by the CG will be very small: (i) we will consider CG 
particle distributions at
scales $\ell_{CG}$ which are large compared to the typical displacements
of particles from their initial lattice position, and (ii) we will
consider CG cells with 
\be \ell_{CG}= \alpha \ell\,, \alpha \in Z^+
\label{def-alpha}
\ee
which are perfectly aligned with the original lattice, and
therefore each contain exactly the same number of points 
of the unperturbed lattice.
To be more  precise in quantifying the importance of these mass 
fluctuations, it will be useful to make the following 
distinction: one can consider the above described
CG as a ``Eulerian'' CG, 
in contrast to a ``Lagrangian'' CG. In 
the latter case we modify step 2 in the above definition
to read:

\begin{itemize}

\item 2'. We ascribe one particle to each cell. The mass, position
and velocity, of this particle are those of the centre of mass
of the points which were, before the application of the initial
random displacements, on lattice sites inside the given cell.


\end{itemize}

In this case --- with CG cells aligned with the
original lattice with $\ell_{CG}$ as in Eq.~(\ref{def-alpha}) ---
the CG particle distribution retains evidently the equal mass 
property of the original distribution.

\subsubsection{Some properties of coarse-grainings}

A few basic results which follow trivially from these definitions
are important to note:

\begin{itemize}

\item A Lagrangian CG of a SL is itself an SL.

\item A Lagrangian CG of a SL with Gaussian PDF is 
itself a SL with Gaussian PDF.

\item An Eulerian CG of a SL converges, in the limit of small displacements,
to the Lagrangian CG on the same grid.

\end{itemize}

The first result is very simple to see. The position of the particles
in the Lagrangian CG can be written as \be {\bf x}_I=
{\bf R}_I + {\bf u} ({\bf R}_I)\,, \ee where \be {\bf R}_I =
\frac{1}{\alpha^3} \sum_i^{(I)} {\bf R}_i \,\,\,\,, {\bf u} ({\bf
R}_I)= \frac{1}{\alpha^3} \sum_i^{(I)} {\bf u} ({\bf R}_i) \ee and the
sums are over the $\alpha^3$ points $i$ whose original lattice sites
are inside the $I$-th cell of the CG lattice.  ${\bf R}_I$ is simply
the center of the new CG lattice cell, and, since the ${\bf u} ({\bf
R}_i)$ are uncorrelated with variance $\Delta^2$, we have \be \av{{\bf
u} ({\bf R}_I) \cdot {\bf u} ({\bf R}_J)}= \delta_K(I,J)
\frac{\Delta^2}{\alpha^3} \equiv \delta_K(I,J) \Delta_{CG}^2.  \ee
Thus the CG particle distribution is a SL, with variance reduced by
${\alpha^3}$.  The PDF of the displacements in the SL can be
calculated given that of the original particle distribution, but in
general has a different functional form. As noted above, however, for
the case of a Gaussian the PDF is the unchanged, because of the
fundamental property of stability of a Gaussian (see, e.g.,
\cite{book}). For the particle distributions we considered here as IC --- SL
with small amplitude Gaussian displacements --- the Eulerian CG is
practically equivalent to the Lagrangian one.

We note one further important property of the CG particle distributions 
defined in this way:  {\it the PS of mass density fluctuations in the original 
particle distribution  and that in the CG constructed from it converge at scales 
larger than} $\sim 2\pi/\ell_{CG}$. This follows from a well-known
argument due to Zeldovich \cite{zeldovich-k4, zeldo, peebles}.
This argument\footnote{We use the word ``argument'' because the heuristic
derivation given by Zeldovich --- and other authors in the cosmological
literature --- fall well short of a proof. See \cite{gabrielli_etal_04} for a 
discussion of this issue and some explicit constructions which 
illustrate the result.}
states that the perturbations to a mass distribution introduced by
moving matter around on a finite scale, while preserving locally the center of 
mass, lead to a modification to the
PS at small $k$ (i.e. smaller than the inverse of the characteristic length
scale) which is proportional to $k^4$. Since, as we have seen above, a
SL has a PS which is proportional to $k^2$ at small $k$, the result follows.

For the case of a CG on a SL, we can in fact derive this result
directly from the expressions Eqs.~(\ref{psSL-gaussian}) and 
(\ref{psSL-leading term}) given above: for 
$\ell_{CG}=\alpha\ell$ we have that $n_0 \rightarrow n_0/\alpha^3$ 
and, we have seen, $\Delta^2 \rightarrow \Delta^2 /\alpha^3$. Thus
the leading $k^2$ term of the PS, given in Eq.~(\ref{psSL-leading term}),
is invariant under the coarse-graining, while the next term,
proportional to $k^4$, is not.

This convergence of the PS of the particle distribution and its CG
quantifies the sense in which the CG leaves invariant 
the large scale properties of the particle distribution. Since a 
continuous fluid-like description of such systems and their dynamics
is expected to be valid for precisely such scales, one would
anticipate that this description should coincide for both
systems. Indeed this is one of the underlying motivations for our
study of this kind of CG.

\section{Coarse-graining of the Initial Conditions}
\label{CG at t=0} 

\begin{table}
\begin{ruledtabular}
\begin{tabular}{cccccccc}
Name & $N^{1/3}$ & $L$ & $\ell$ & $\Delta$ & $\delta$ & $m/m_{64}$ \\ 
\hline 
SL64     & 64 & 1& 0.015625& 0.00195   & 0.125    & 1    \\ 
CG32$_0$ & 32 & 1&0.03125  & 0.00069   & 0.0221   & 8    \\ 
CG16$_0$ & 16 & 1&0.0625   & 0.00024   & 0.0039   & 64   \\ 
\hline 
\end{tabular}
\end{ruledtabular}
\caption{
Details of the three sets of initial conditions studied in this
section. SL64 is the original (or ``mother'') particle distribution  
and CG32$_0$ and CG16$_0$ are the two ``daughter'' particle distributions 
obtained by CG it as described in the text. 
$N$ is the number of particles and $L$ is the box size
(arbitrarily set to unity). The mass $m$ of the particles is given 
in units of that in SL64, i.e., $m_{64}$. The mass density is of
course the same by construction in all the particle distributions,
$\rho_0 = Nm/L^3=1$. 
\label{tab:sl-summary} }
\end{table}

In this section we study the evolution from a SL initial condition,
and from CGs defined on this initial particle distribution. We first
present and analyze our numerical results, and then analyze what can
be learned from them about the non-linear clustering in these
systems. Specifically we will see that the non-linear two-point
correlation function, which develops in a given such system, can be
well reproduced, over a significant range of time and spatial scale,
in a CG of the system evolving in a regime in which the relevant
dynamics are driven by two body interactions of NN pairs.  There is,
however, a small discrepancy in the amplitude of the correlation
function which we show can be quantitatively understood using PLT (and
which is an effect of discreteness, i.e., due to the different
particle density in the simulations).  We then show that these, and
other, differences between the {\em mother} (i.e. the original
particle system) and {\em daughter} (i.e. the CG particle
distribution) systems will increase, without limit, as the parameter
$\alpha=\l_{CG}/\ell$ does. This then leads us naturally to the CG
presented in the following section.

As our original (``mother'') system we take here, and in
the next section, the following IC: a simple cubic lattice 
with $N=64^3$ particles, to which we apply a shuffling
with Gaussian PDF and normalized shuffling parameter
$\delta=0.125$, i.e., as given by Eqs.~(\ref{eq:pu})
and (\ref{eq:defDelta}) with $\Delta=0.125 \ell$ 
(and $\ell=L/N^{1/3}$). We will call this initial condition 
SL64. 

We consider then also two other IC which are Lagrangian
CG as defined above: CG32$_0$ and
CG16$_0$, defined respectively on the $32^2$ and
$16^3$ sub-lattice of the original lattice (i.e.
with $\alpha=2$ and $\alpha=4$) respectively. Since 
$\delta$ is small this CG is, as we have
noted above, in practice equivalent to the
Eulerian CG\footnote{The probability $p_{out}$ that any
given particle of the initial $N=64^3$ lattice is displaced
by its shuffling so that it falls outside the CG cell of
the $32^3$ CG can be written as
$p_{out} =1-[\sqrt{3/2\pi} \int_{-1/2\delta}^{3/2\delta}
e^{-3x^2/2} dx]^3$. For $\delta=0.125$ this corresponds
to $p_{out} \sim 10^{-12}$. Thus $Np_{out} \ll 1$.}.
Given that $\ell_{CG}=\alpha \ell$ and 
$\Delta_{CG}^2=\Delta^2/\alpha^3$, we have
\be
\delta_{CG} \equiv  \frac{\Delta_{CG}}{\ell_{CG}} = \frac{\delta}{\alpha^{5/2}}
\label{scaling-delta}
\ee
which explains the values of $\delta$ which appear in
Table \ref{tab:sl-summary}.  Further, in order to keep the mass
density $\rho_0$ fixed, we take
\be
m_{CG}= \frac{m}{\alpha^{3}}\,.
\label{scaling-mass}
\ee where $m$, the mass of particles in the original SL, is
(arbitrarily) normalized to unity in SL64.

The initial velocities of SL64 are fixed using Eq.~(\ref{vel-IC}).
In the two CGs they are
fixed, as described above, by the condition that the CG
points are ascribed the velocity of the center of mass 
of the set of points they represent. It is evident that this is
in fact equivalent (for the Lagrangian CG)
to also using Eq.~(\ref{vel-IC}) on the CG points
directly.

\begin{figure*}
\caption{Snapshots at times $0$ (top line), $3$, $6$ and $8$ (bottom line)
  of SL64 (left column), CG32$_0$ and CG16$_0$ (right column). 
  What is shown is a projection on the
  $x$--$y$ plane of a slice of thickness 0.6 along the $z$ axis.
\label{fig:snapCGt0} }
\end{figure*}

\subsection{Results of numerical simulations}
\label{Results of numerical simulations}

\begin{figure*}
\includegraphics[width=\textwidth]{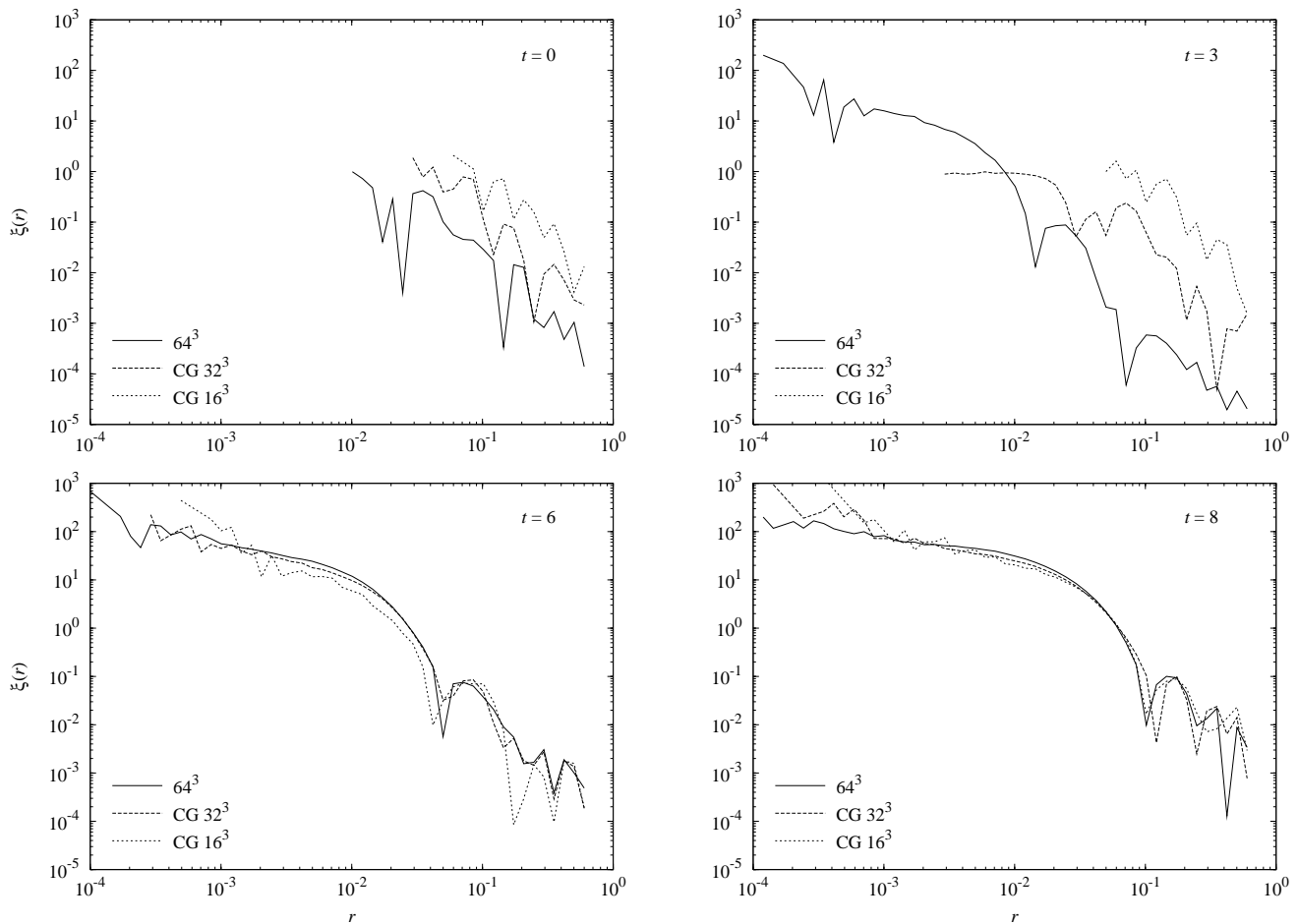}
\caption{Evolution of $\xi(r)$ in SL64 and in
CG32$_0$ and CG16$_0$. The times are given in 
the plots. 
\label{fig:CGt0xi}}
\end{figure*}

The numerical code used for our simulations is, as in \cite{sl1}
and \cite{marcos_06}, the publically available GADGET code 
\cite{gadget}. It uses a tree algorithm for the calculation
of the gravitational force. The singularity at $r=0$ in the latter is
regularized, for numerical efficiency, at very small scales, i.e.,
well below the initial inter-particle spacing $\ell$. The simulations
here (as in \cite{sl1}) use a ``smoothing parameter''
$\varepsilon=1/15 \cdot 1/64 \approx 0.0010$ in the units used in
Table \ref{tab:sl-summary}, i.e., about one fifteenth of the smallest
{\it initial average inter-particle distance} in our simulations
\footnote{For this code \cite{gadget} this means that the 
force is the exact gravitational force for separations
greater than $2\varepsilon$. For the precise functional form 
of the regularization below this scale, see \cite{gadget}.}.
We assume here, as in \cite{sl1} and indeed generically
in all such cosmological simulations, that the results
we consider as physical, starting from separations a 
few times larger than $\varepsilon$, are essentially 
independent of this choice of $\varepsilon$. As 
in \cite{sl1} we have tested this assumption, and found 
it to be good, by
resimulating identical initial conditions changing
only $\varepsilon$ to considerably smaller values.

\begin{figure}
\includegraphics[width=0.5\textwidth]{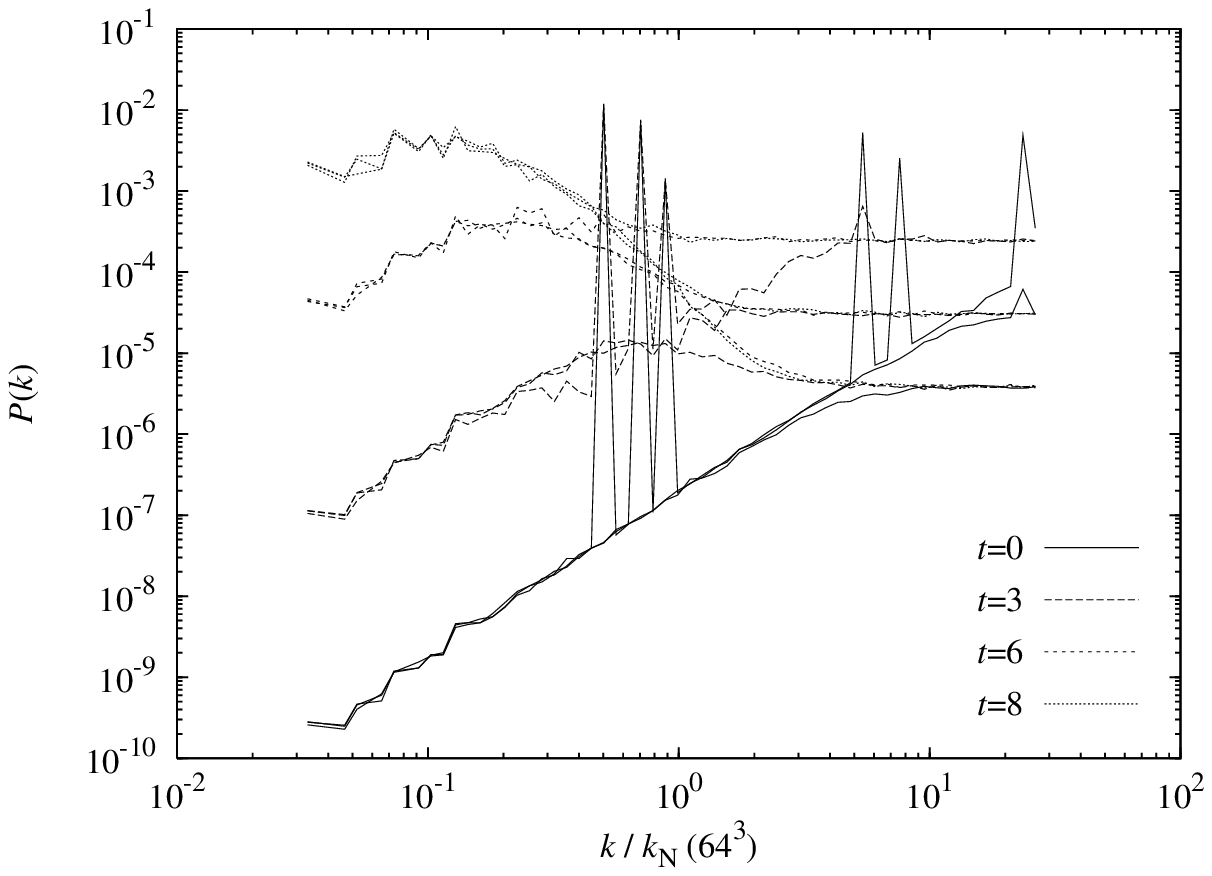}
\caption{Evolution of the PS in SL64, CG32$_0$ and CG16$_0$.
Times are as indicated in the legend. The $x$-axis is normalized 
to $k_N$ in SL64.  At each time the different systems can be 
most easily distinguished by looking at the PS at large $k$,
which approaches cowards $1/n_0$. The peaks which are evident in
in the PS are the Bragg peaks of the lattice, which progressively
disappear as the lattice structure is washed out by the evolution.}
\label{fig:CGt0ps}
\end{figure}

In Fig.~\ref{fig:snapCGt0} 
are shown snapshots of two-dimensional
projections of the simulations starting from SL64, 
CG32$_0$ and CG16$_0$. Here, and in the rest of this paper, 
{\it time is given in units of dynamical time} $\tdyn$. 
We stop our analysis at $t=8$ as for longer times finite size 
effects begin to be important as the size of the structures 
formed approach  the size of the box. As time evolves 
structures first form in each simulation at small scales
--- well below the initial grid size $\ell$ --- and 
then subsequently at larger scales. Clustering develops
first in SL64, and only later in its CG.
While no similarity in the fluctuations is {\it visually}
discernible when the particle distributions are grid-like, in the
evolved configurations one sees clearly that the structures
which formed in the CGs trace approximately
those in the ``mother'' distribution. 

In Fig.~\ref{fig:CGt0xi}  we show the evolution of the two-point 
correlation function $\xi(r)$ at the times indicated.
These plots reflect in a quantitative manner the information
gleaned from the visual inspection of the snapshots in the
previous figure. The three systems show initially no discernible
similarity: the real space correlation function is in each
case dominated by the features of the different underlying
lattice. Strong clustering (with $\xi \gg 1$) develops 
first at small scales in the SL64 simulation, and then
subsequently at larger scales. When the coarse-grained
systems start to evolve a clustered regime, their correlation
functions start rapidly to resemble strongly those in
the ``mother'' system. We note, however, that there is
a transient time in which the coarser system has 
a correlation function resembling that of the ``mother
system'', but of a slightly lower amplitude. This ``lag''
of the coarser system is here manifest in the plot at 
$t=6$. We will return to this point below.

In Fig.~\ref{fig:CGt0ps} is shown the evolution of the PS in the three
simulations at different times steps, with $k$ in units of the Nyquist
frequency $k_N=\pi/\ell$ of SL64.  The three curves at $t=0$
correspond to the theoretical PS given in
Eq.~(\ref{eq:exactPS}), with the appropriate Gaussian form
$\tilde{p}(k)$ and value of the shuffling variance. 
The agreement of the PS at small $k$ discussed above is
clearly seen (and one can verify the accuracy of the $k^2$
behavior). The structure of peaks, which is different 
in each initial condition, comes from the second term 
in Eq.~(\ref{eq:exactPS})\footnote{Note that not all the Bragg
peaks in the PS [cf. second term in Eq.~(\ref{eq:exactPS})] are
visible in the plot.) Indeed the first Bragg peak in SL64 is at
$k=2k_N$, while the first one in CG32$_0$ and CG16$_0$ are at $k=4k_N$
and $k=8k_N$ respectively. The reason is that the PS is calculated
using a sampling of $\ve k$, and the peaks are so narrow that they are
missed. 
}.
Note that while the differences between the distributions at the
initial time manifest themselves in the PS only at large $k$, 
we have seen that the initial correlation functions are very
different at all scales\footnote{This ``localisation'' of
the discrete characteristics of the distributions in reciprocal
space, and ``delocalisation'' in real space, is also a feature of 
the initial conditions of cosmological simulations.
See \cite{michael_bruno} for a detailed discussion.}. At $t=3$ this 
peak structure is still visible only in
the CG16$_0$ simulation, corresponding to the fact that the lattice
structure is still present in this simulation which has evolved little
non-linear clustering at this time. The relative difference between
two other simulations is now less than at the initial time, showing
again that once structures develop in the ``daughter'' particle
distribution they trace those in the original distribution quite
accurately. At later times this behavior is also seen for the
CG16$_0$ simulation, as it ``catches up'' with the other ones.

These observations are, to a first approximation, very much in line 
with the qualitative picture of the evolution of clustering widely 
accepted in the theory of structure formation perturbations in 
cosmology (see, e.g., \cite{peebles}, 
or \cite{padmanabhan_short} for a concise summary): 
non-linear gravitational evolution transfers 
power from larger to smaller scales (by collapse) so efficiently that, 
for initial PS with $n<4$ (as is the case here, $n=2$), the clustering 
amplitudes in the non-linear regime at a given time depend essentially 
only on the initial power in modes corresponding to larger scales. Thus what
is relevant to recover the non-linear evolution of the system is to
include this initial power at larger scales\footnote{See, e.g., 
\cite{Little+weinberg+park_1991} for a numerical study of this 
issue in cosmological simulation, and \cite{peacock} for a
widely used phenomenological ansatz for ``reconstruction'' of
the non-linear PS from the initial conditions.}.  The coarse-grainings 
we have applied conserve, by construction,  the initial power at 
larger scales, and thus can begin to reproduce clustering correctly 
at scales smaller than the CG scale once the smallest scales with the 
``correct'' power go non-linear. More qualitatively,
the approximate criterion used in cosmology to determine when
non-linear evolution should set in, adapted to the present case
of a static universe, is
\be
k^{3} P(k) e^{2t/\tdyn} \sim 1\,,
\label{collapse-criterion-cosmo}
\ee 
i.e., the fluctuations of wavenumber $k$ satisfying this relation
at a given time are those ``going non-linear'' according to the
linearized fluid theory at that time
\footnote{
The quantity $k^3 P(k)$ is equal, up to a numerical factor, to
the variance of mass fluctuations averaged over a spherical Gaussian
window function of radius $\approx 2\pi/k$ cf. \cite{peebles}. We have
adapted the criterion here to the case of a non-expanding universe,
by using the linear theory growth factor appropriate for this case.}.
Assuming that the coarse-graining procedure leaves only
initial fluctuations in reciprocal space below the Nyquist 
frequency, the largest $k$ available to go linear, which is
therefore the first one to do so (since $P(k) \propto k^n$
with $n > -3$), is the Nyquist 
frequency, $k_N=2\pi/\ell_{CG}$.  The latter decreases, and 
therefore non-linearity sets in later, as $\ell_{CG}$ increases. 
To determine the time scale this argument predicts, it is 
convenient to note that,  using Eqs.~(\ref{psSL-leading term})
and (\ref{scaling-delta}), the criterion 
Eq.~(\ref{collapse-criterion-cosmo}) with
$k=k_N$, may be rewritten  
\be
\delta_{CG}^2 e^{2t/\tdyn} \sim 1\,.
\label{fluid-collapse-CG}
\ee
Thus the time scale predicted for the first non-linear structures
is just the time of fall of neighbouring particles
on one another, {\it calculated in the fluid limit} (i.e. using 
the ZA as in Eq.~\ref{ZA-grav-field-1}.). We will return to
the question of the accuracy of this prediction below, and
explain that it will be a good approximation if discreteness 
effects up to this time (i.e. when particles first approach
one another) are small. For the numerical simulations here 
this is indeed a reasonable first approximation, because
the duration of this period is not too long (at most a few
dynamical times).

To the  extent that the evolution of  the measured two point 
quantities  agrees (after some time),  one can conclude that  
this evolution can  depend only on quantities which are the same 
in all the systems. Specifically, as we have just discussed,  
it can (and does) depend on the  amplitude of  the  
initial fluctuations   at the relevant  larger scales, but 
it {\it  does not depend  on  the particle density}.  From
this latter fact one might be tempted to conclude that the dynamics of
the system  should  be well  described   by one in which  the  density
fields, for example,  are smooth functions, i.e.,  by a description in
which the discrete nature of the system is of no relevance\footnote{We
have in mind, specifically,   the description  of  the dynamics  by  a
Vlasov-Poisson  system  of equations. In cosmology the aim of N-body
simulations   is to reproduce  the   evolution   of  this system    of
equations.  We  will discuss  this  question in greater detail  in the
conclusions section.}. Independence  of measurable  quantities of  the
particle   density    does   not,     however,   establish   such    a
conclusion.  Indeed  we   will see   now  that,   on the  contrary,  a
description     of the  dynamics    in   a    manifestly particle-like
(i.e. discrete) way   captures  essential aspects of the    evolution.

We consider now more closely both the differences and 
similarities between the evolved simulations. We note in 
particular that the CG systems trace very well the two
point correlations in the original system, already
at a time when a very simple model for the development 
of these correlations, in terms of two body interactions,
is valid. There is, however, a small discrepancy in
the amplitudes of the correlation functions, which
we explain.  This leads us naturally to consider 
the extrapolation of these results to the case that the 
CG scale becomes very large compared to the inter-particle 
distance in the original particle distribution.

\subsection{Non-linear correlations at early times}

In the results above we have seen clear evidence for the
convergence of the correlation properties of the 
evolved coarse-grained IC toward those of the
evolved ``mother'' particle distribution. There is, however, a 
range of time scales in each case in which there are
already significant non-linear correlations in the 
CG simulation, but the agreement, notably
in the amplitude of the correlation function, is not so 
good.  This discrepancy can be understood by analyzing
the dynamics giving rise to these correlations at early
times.

\begin{figure}
\includegraphics[width=0.5\textwidth]{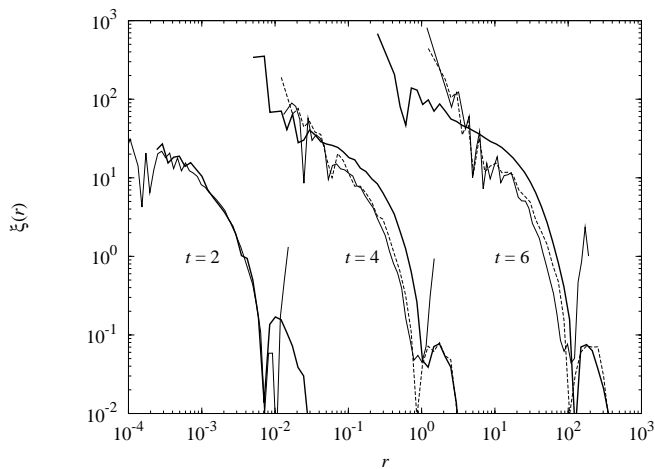}
\caption{Reconstruction of $\xi(r)$ using the NN distribution
  $\omega(r)$ in Eq.~(\ref{omega1}). At time $t=2$, it is applied 
to SL64: the thick line represents the directly measured $\xi(r)$,
while the thin line is the reconstruction with $\omega(r)$. 
The chosen times $t=4$ and $t=6$ for the other two plots correspond 
to times a little before the correlation function of CG32$_0$ and 
CG16$_0$ catch up with SL64. In the two cases, we have plotted
$\xi(r)$ in SL64 (thick lines), $\xi(r)$ in the CG system (dashed
line) and its reconstruction with $\omega(r)$ (thin continuous
line).  This allows one to see that at these times, when the first 
non-linear correlations are created in all these systems, they are 
well accounted for purely by correlation of NN pairs, which develop
through the interaction of such pairs. For clarity, the data at the
two later times have been rescaled along the
  $x$-axis (by a factor $50$ and $50^2$ respectively).}
\label{NNreconstruction}
\end{figure}

We have shown in \cite{marcos_06} that the evolution at early times
--- when the displacements are small compared to the lattice spacing
--- is very well described by the perturbative approach (PLT)
described briefly in Sect.~\ref{PLT} above.  It breaks down when pairs
of particles start to approach one another. In the immediately
subsequent phase strong correlations (with $\xi \gg 1$) begin to
develop at small scales, and it is natural to hypothesize that an
important role is played at this time by the interaction of particles
with their NN.  A simple way to test for this is, as explained in
Sect.~\ref{characterisation} above, is to see whether the relation
Eq.~(\ref{omega1}) holds. To do this we ``reconstruct'' the two-point
correlation function $\xi(r)$ from the measured NN distribution using
this relation, and compare with the measured $\xi(r)$ at each time.
The results are shown in Fig.~\ref{NNreconstruction}. Inspection
shows that in each of the simulations the relation holds to a very
good approximation. The times at which we have chosen to plot the 
comparison in each case is the largest time at which the 
agreement is observed to be good. This time increases as the IC 
becomes coarser because, as discussed above, the phase in which
PLT is valid increases as the CG scale does. Therefore the time of
switching on of binary collapse becomes correspondingly larger.

The CG systems thus develop significant non-linear
correlation in the phase in which the relation Eq.~(\ref{omega1}) 
holds. We infer that the correlation measured is essentially
the result of two body correlations developing through the
interaction of particles with their NN. What is very striking,
as it has been pointed out in \cite{Baertschiger:2004tx, sl1},
is that the two-point correlation function traces very well, at
least {\it in its shape}, that in the mother simulation, even
though in the latter the dynamics are at this time no 
longer NN dominated.  At these times in fact the mother simulation
is evolving in the self-similar manner described in 
\cite{sl1}, for which the temporal scaling [i.e. the dynamical
exponent given in Eq.~(\ref{eq:self-sim-xi})] can be determined from a
fluid-like description of the system\footnote{We recall that the
scaling is inferred from that of the large scale fluctuations in the
model treated in the linearized fluid limit, cf. \cite{sl1}.}. Thus
the complex many body dynamics of the mother system (whether mean
field or not) is well approximated, at the relevant scales, as
particle-like ``blobs''  each moving under the effect of the
single nearest ``blob''.

\subsection{Extrapolation to large $\ell_{CG}/\ell$}

There is, however, as we have noted also in the discussion
of Fig.~\ref{fig:CGt0xi}, a small (but discernible) discrepancy 
between the {\it amplitude} of the correlation function in
the CG system and that in the original one. 
The CG systems thus appear to lag behind for a time,
before ``catching up'' subsequently with the mother system.
Thus the dynamics of the formation of the correlation 
in the mother system, in the asymptotic self-similar regime,
appears to be very well represented by that of CG particles, 
but these particles fall toward one another by NN 
interactions slightly slower than the particle-like 
``blobs'' of the original particle distribution. 

This difference can in fact be understood from the 
perturbative approach (PLT) of \cite{marcos_06}, which 
we outlined above in Sec.~\ref{PLT}. For most modes
of the displacement field applied to a simple cubic 
lattice, the largest eigenvalues characterizing  
growth  are in fact {\it smaller} than the (single)
exponent characterizing the fluid limit, given 
by the ZA as in Eq.~(\ref{ZA-grav-field-3}) for
our initial conditions. Thus the evolution
of displacements is, typically, {\it slower} in the discrete
particle distribution (described accurately by PLT)
than in the fluid limit\footnote{We write ``typically'' because there are
a very small fraction of modes, in a simple cubic lattice,
which grow slightly {\it faster} than in the fluid limit.
Given their small number, and the small difference in
the growth rate, they are significant in the average
only at {\it very} long times.}. Further the discrepancy in any given quantity 
in the two treatments becomes more important as the time 
increases (so long as PLT itself remains valid). Specifically,
for example, the criterion Eq.~(\ref{fluid-collapse-CG}), 
derived above in the fluid (ZA) limit, for the time of onset 
of the phase of development of strong non-linear correlations 
(dominated, at first, as we have seen by NN interactions) 
can be calculated more accurately using PLT. The result 
can be written schematically as
\be
\delta_{CG}^2 e^{(1-\eta)2t/\tdyn} \sim 1\,.
\label{fluid-collapse-CG-PLT}
\ee
where $\eta$ is different from zero, and typically 
positive\footnote{In \cite{marcos_06} the full gravity, 
PLT and fluid evolution is calculated for the average 
relative displacement between particles initially on 
neighboring sites of a SL configuration, and
the ``lag'' effect, described approximately by
Eq.~(\ref{fluid-collapse-CG-PLT}) with $\eta>0$,
is manifest (see e.g. Fig. 8 of \cite{marcos_06}).
At {\it very} long times the sign of the
effective $\eta$ will change as the few modes
which grow more rapidly than in the fluid (see preceeding 
footnote) come to dominate.}. Thus the discrepancy between the 
fluid prediction of Eq.~(\ref{fluid-collapse-CG}) 
and that of PLT Eq.~(\ref{fluid-collapse-CG-PLT}) will 
increase as $\delta_{CG}$ decreases, i.e.,
as $\alpha$ increases [cf.Eq.~(\ref{scaling-delta})]. 

It follows that if we extrapolate the coarse-graining to
arbitrarily large $\alpha$, an arbitrarily large difference 
in the evolution of the CG daughter and that of the
original distribution will result\footnote{Indeed for 
asymptotically large $\alpha$ one will
obtain completely different qualitative behaviours of
the two systems, as the small number of eigenmodes 
which grow faster than in the fluid limit will dominate.
The eigenmode with the very largest eigenvalue
(about ten percent larger than the fluid one) corresponds 
to adjacent pairs of planes of the lattice falling toward one 
another \cite{marcos_06}. In the limit that $\delta$ is sufficiently
small that such a mode can collapse before any other 
one, there will be an imprinted long range order in
the structures formed which is sheerly an artifact
of the CG lattice.}. Such behavior could, in priciple,
be observed, of course, in numerical simulations 
like those performed here, if the initial $\delta$ 
is sufficiently small. In practice this is not 
numerically feasible as it requires unattainable
accuracy on the force\footnote{In \cite{marcos_06}, 
however, related oscillating modes of the lattice
have been simulated using a simplified code exploiting 
the symmetries of certain modes.}. We note also that the small
time-lag effect which we have identified here in our simulations 
was not seen in the results of  \cite{Baertschiger:2004tx}, 
because the SL simulations considered started from larger
values of $\delta$ (larger, specifically than that
in CG16$_0$, cf. Table~\ref{tab:sl-summary}).
 

\section{Coarse Graining at a finite time}
\label{sec:later_time_evol}

We have seen in the previous section that the CG
on a grid of the SL initial conditions, i.e., of the system at the
initial time, leads to an evolution which diverges further and
further from that of the original system as the ratio of
the CG scale to the scale of the original grid
increases. The reason is that the CG system of
particles evolves for a longer time in the regime in which
discreteness effects induced by the lattice accumulate. 

In this section we define and study a different CG which, by
construction, avoids this problem: instead of coarse-graining the
initial condition we coarse-grain the original system {\it when it has
evolved for a finite time}.  The idea is simple: we choose the time at
which we coarse-grain the system at a given scale by a prescription
which means that, approximately (in the sense discussed in what
follows), the parameter $\delta_{CG}$ (i.e. the one point variance of
the displacements of the CG particles off their lattice) is {\it equal} to
that of the original particle distribution. The CG system will therefore
spend approximately the same amount of time in the regime of validity 
of PLT as the original system.

\subsection{Definition of coarse-graining at a finite time}

As in the previous section we consider coarse-graining a
particle distribution with lattice spacing $\ell$ using a CG grid
which is also a lattice with lattice spacing  
$\ell_{CG}= \alpha \ell$, where $\alpha$ is an integer. 
The prescription we use for the time $t_{CG}$ at which
we perform the coarse-graining is
\begin{equation}
t_{CG}= \tdyn \frac{5}{2} \ln \alpha \,.
\label{prescription-CGT}
\end{equation}

The reason for the choice Eq.~(\ref{prescription-CGT}) is simple :
{\it if} the Zeldovich approximation is assumed to be 
exact at the scale of the coarse-graining, the CG system is
then exactly a SL {\it with the
same value of the normalized shuffling parameter} $\delta$.
Thus the CG at time $t_{CG}$ is related to the original
system at $t=0$ by a simple rescaling of all length 
scales\footnote{This is for the infinite lattice and exact gravity. 
It is thus implicit here that we consider the regime where 
finite box size effects, and effects depending on the finite 
softening $\varepsilon$ may be neglected.}.
Let us see that this is the case.

When the ZA is valid, in the form as given in Eq.~(\ref{ZA-grav-field-1}), 
it follows that all particle displacements are simply amplified 
in time by the factor $e^{t/\tdyn}$. Thus a SL remains a SL, with 
a growing variance of its random uncorrelated displacements. If
we perform a Lagrangian CG at time $t$
we therefore obtain a SL with normalized shuffling $\delta_{CG}$ 
given as Eq.~(\ref{scaling-delta}), i.e., 
\be
\delta_{CG} = \frac{\delta(t)}{\alpha^{5/2}}= \frac{\delta(0)e^{t/\tdyn}}{\alpha^{5/2}}\,.
\label{scaling-delta-evolved}
\ee Taking the prescription Eq.~(\ref{prescription-CGT}) for the time
at which the CG is performed, we thus have $\delta_{CG}=\delta(0)$.
Further we recall that in the ZA the velocities scale in proportion to
the displacement [cf. Eqs.~(\ref{ZA-grav-field-1}) and
Eq.~(\ref{ZA-grav-field-2})]. Thus the velocities in the CG particle
distribution also simply rescale those in the original distribution in
proportion to $\alpha$. The CG system is thus (statistically)
identical to the original system up to a rescaling of spatial
variables of the original system ${\vec x} \rightarrow \alpha {\vec
x}$~\footnote{We assume as the previous section that we have a
Gaussian PDF.}.

In reality of course the ZA is not exact. Above we have assumed in
fact that it is exact for the evolution {\it of each particle} of the
mother distribution until an arbitrary time (since $t_{CG}$ increases
arbitrarily as $\ell_{CG}$ does). Clearly this is not even a good
approximation in general. We need, however, only to make a much weaker
assumption to recover the result (that the {\it finite-time} CG is a
rescaling of the mother SL): all we require in fact is that {\it the
center of mass of the CG cell moves according to the ZA until
$t_{CG}$}. Thus, specifically, we require only that the ZA be a good
approximation when we neglect the gravitational forces acting between
particles within the same CG cell (since such forces do not affect the
center of mass motion).

We have seen that the ZA can be derived as an appropriate limit of the
full particle evolution using PLT. This latter treatment is of course
only valid until the time at which particles approach one another for
the first time. We can, however, modify it to understand the nature of
the approximation required here. To do so, we break the force on a
particle into two parts: that due to particles in a region of size
$\eta \ll \ell_{CG}$ about it, and the residual force from particles
outside this region. The local part of the force will have a
negligible effect on the motion of the center of mass of the regions
of size $\ell_{CG} \gg \eta$, and so we can neglect it. The second
piece of the force can be linearized in the relative displacements as
in PLT, which now remains until particles move a distance of order
$\eta$. Because the dynamical matrix has been modified only up to a
finite scale, the diagonalisation which gives the eigenvectors and
eigenvalues for the displacement field will converge to the same
behavior at small $k$ as in the full PLT. The ZA will therefore still
be a valid approximation in this limit, with corrections at any finite
$k$ which can in principle be calculated. Thus, in summary, we expect
the ZA to be a good approximation as required, if the scale of
non-linearity is smaller than $\ell_{CG}$ at time
$t_{CG}$\footnote{This will necessarily be the case here as the
$\delta$ of the original particle distribution, which controls the
time at which any scale goes non-linear, has been chosen small
compared to unity (but not too small, as discussed above).}. We also
expect, as in the previous case of CG at the initial time, that, at a
given length scale, the deviations from ZA will diverge as the time
for evolution with this ``modified PLT'' description does.  We
therefore expect the result to break down if the initial $\delta$ in
the mother particle distribution (and therefore that in the CG
distribution also, since they are equal) becomes itself arbitrarily
small.

The CG at a finite time, defined as we have just discussed, is useful 
with respect to  understanding the ``self-similar''
properties observed in the evolution of a SL, extensively
discussed in \cite{sl1} (and summarized in 
Sect.~\ref{Gravitational clustering in a Shuffled Lattice}
above). In the case that the CG is just a rescaling of the
initial SL (i.e. in the approximation that the ZA is exact), 
it follows that 
\begin{equation}
\xi_{CG} (r, t_{CG} +t) 
= \xi \left(\frac{r}{\alpha}, t\right) = \xi (r e^{-2t_{CG}/5\tdyn} , t)
\label{self-sim}
\end{equation}
where $\xi_{CG}$ is the correlation function measured in the evolved CG
system and $\xi(r)$ that in the original one. Now, if the CG system
indeed reproduces the evolution of the original system, i.e.,
$\xi_{CG} (r, t) = \xi (r,t)$ for $t>t_{CG}$, the self-similarity
relation Eq.~(\ref{eq:self-sim-xi}) follows.

\subsection{Numerical simulations}

Our original (``mother'') system is the same SL64 initial condition
considered in the previous section. Our coarse-grainings
are now performed at the same length scales, i.e., $\alpha=2$
and $\alpha=4=2^2$. Following the prescription just given we
therefore coarse-grain the original particle distribution evolved
to times $t_{CG}$ and $2t_{CG}$ respectively, where 
 \begin{equation}
t_{\rm CG}= \frac{5}{2} (\ln 2) \tdyn \approx 1.73 \tdyn \,.
\label{tCG}
\end{equation}
Unlike the case of the CG at $t=0$, the Eulerian
and Lagrangian CGs are in each case not the same.
We thus define for each case two CGs so that
we have four in total CG32$_t$, CG32$_t^\prime$, CG16$_t$
and CG16$_t^\prime$. The unprimed acronyms refer to the Eulerian
(non-equal mass) simulations, the primed ones to the
Lagrangian (equal mass) simulations.  Since the IC are actually 
such that particles have moved only a small distance compared 
to $\ell_{CG}$ at the time of the corresponding coarse-graining, 
the differences should not be very important, as they come
only from particles close to the surface of the CG cell being
re-attributed to neighboring cells.

\begin{figure}
  \includegraphics[width=0.45\textwidth]{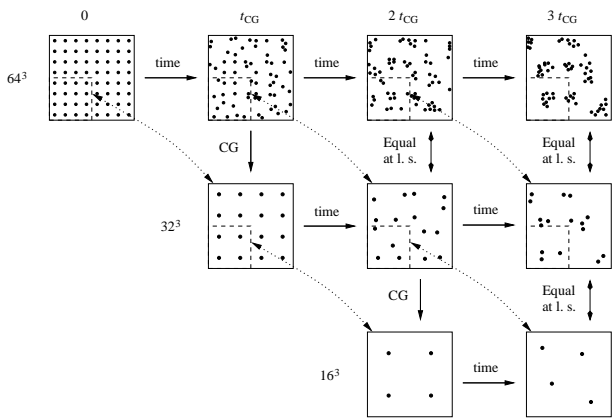}
  \caption{Schematic representation of the ``CG in time''
  experiment. The $64^3$, $32^3$ and $16^3$ systems are represented
  from top to bottom, and the time evolution is represented in steps
  of $t_\text{CG}$ from left to right. The arrows marked ``CG''
  indicate that a CG is done (note that in the experiment done, the CG
  for the $16^3$ system is obtained from the $64^3$ system and not
  from the $32^3$). The double arrows marked ``l. s.'' {\it (large
  scales)} indicate that the system should share the same fluctuations
  at sufficiently large scales. The dashed arrows and the dashed
  squares illustrate the ``self-similarity'' argument: a square
  defined (partially or wholly) by the dashed lines are supposed to be
  identical up to a rescaling of length scales to the system to which
  the dashed arrow points.
\label{fig:scheme_total_CGintime}
  }
\end{figure}

\begin{figure*}
\caption{Snapshots of the evolution of SL64 (left column), CG32$_0$
  and CG32$'_t$ (right column) at times 2,4,6 and 8. The points shown 
are obtained by projection on the
  $x$--$y$ plane of a slice of thickness 0.4 along the $z$ axis.
\label{fig:snapcompare}
 }
\end{figure*}
In Fig.~\ref{fig:scheme_total_CGintime} we give  a schematic
representation of this CG scheme.
In Fig.~\ref{fig:snapcompare} 
are shown snapshots in time of the evolved 
SL64, CG32$_0$ and CG32$_t^\prime$ systems starting from a time $t=2$ just a 
little after $t_{CG} \approx 1.73$ at which the latter distribution is 
defined. CG32$_0$ and CG32$_t^\prime$ are clearly very similar, but visually 
one can discern that CG32$_t^\prime$ traces slightly better than CG32$_0$ the 
morphology and clumpiness of the structures in the mother particle distribution.
\begin{figure*}
  \includegraphics[width=\textwidth]{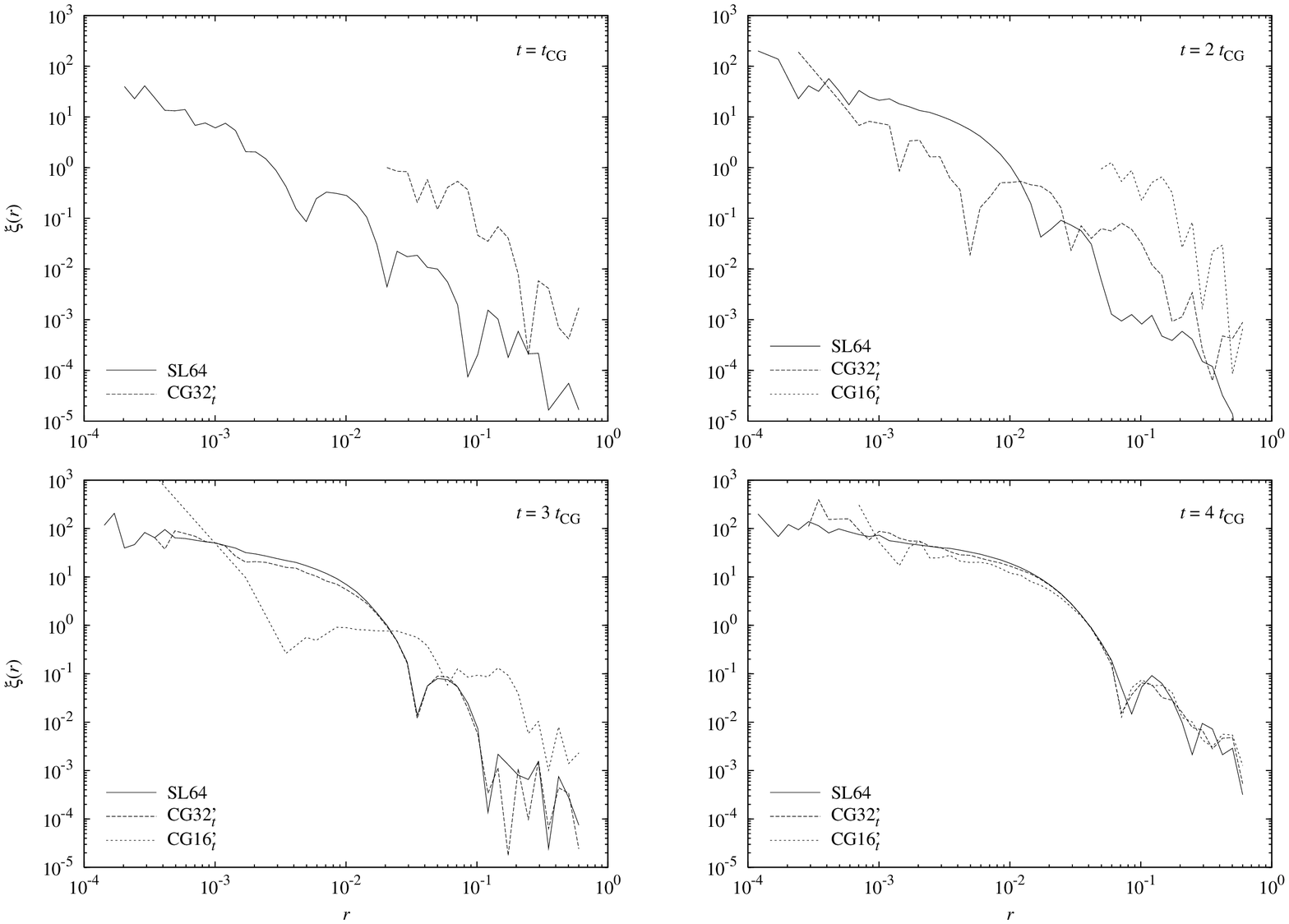}
  \caption{Evolution of $\xi(r)$ in SL64 and the Lagrangian (equal
  mass) coarse-grainings CG32$_t^\prime$ and CG16$_t^\prime$. The
  different plots correspond to the times 1, 2, 3 and 4 in units of
  $t_\text{CG}$.
\label{fig:CGintimeXiCompare}
  }
\end{figure*}

\begin{figure*}
  \includegraphics[width=\textwidth]{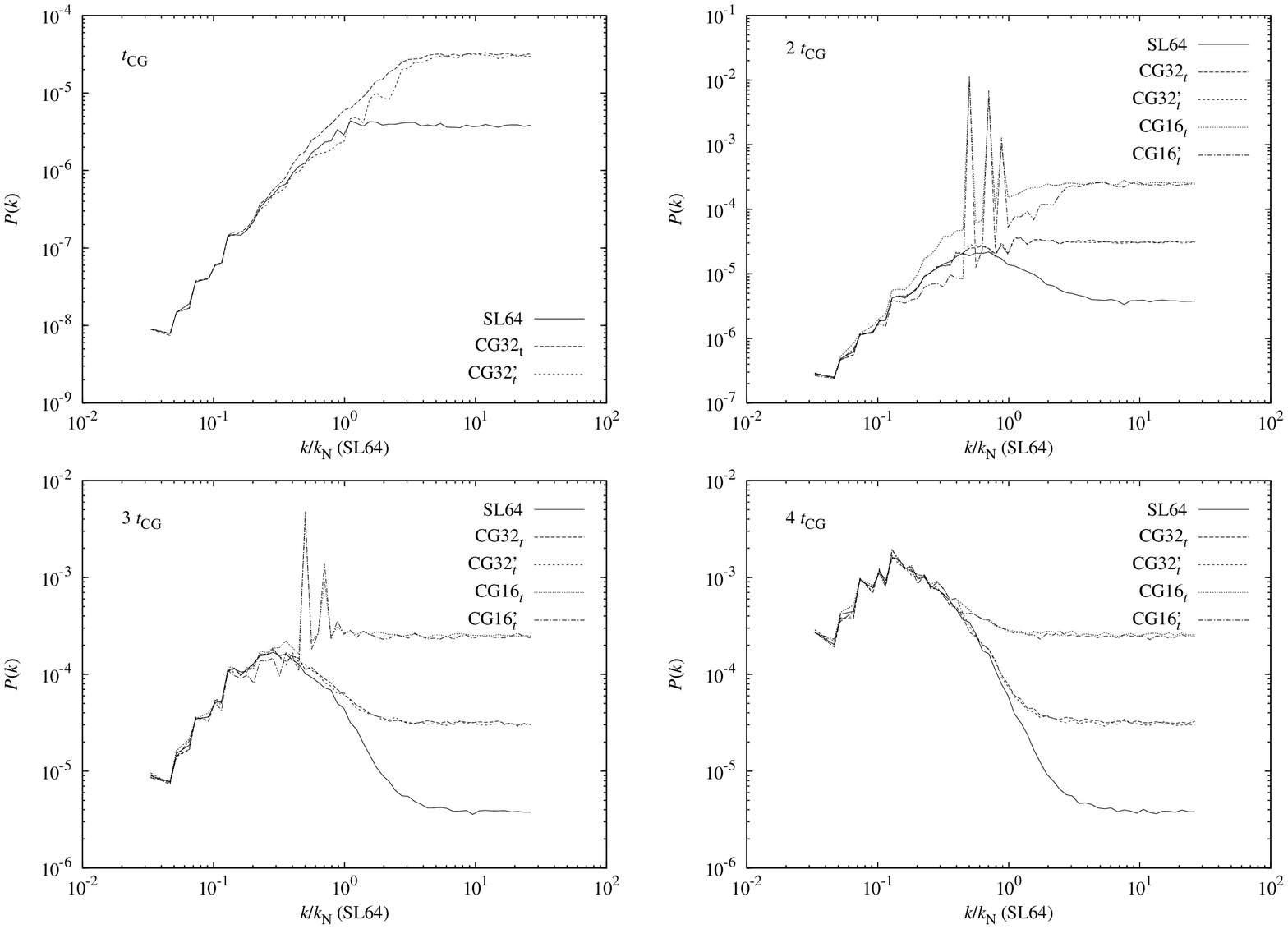}
\caption{Evolution of the PS in SL64 and the different CG in time (see
the legend). The different plots correspond to the times
1, 2, 3 and 4 in units of $t_\text{CG}$. \label{fig:CGtimePSCompare}}
\end{figure*}

In Fig.~\ref{fig:CGintimeXiCompare} is shown the evolution
of the two-point correlation function in SL64, CG32$_t^\prime$ and 
CG16$_t^\prime$. We do not show here results for the Eulerian CGs 
because the (number-number) correlation function we can estimate 
straightforwardly is not equal in this case to the density-density 
correlation function (which is the quantity we are interested in).
As in the previous section we see that the CG particle distributions begin
to develop non-linear correlations later than in the mother 
distribution, but that the relative difference between the
correlations in the two particle distributions becomes negligible 
subsequently.
In Fig. \ref{fig:CGtimePSCompare} is shown the temporal
evolution of the PS in the different particle distributions. In this
case we show both the Lagrangian and Eulerian CG 
distributions as our estimator of the PS is indeed
that of the mass density fluctuations [i.e. it takes
into account the different masses of the particles,
cf. Eq~(\ref{eq:deltak})]. 
We see again qualitatively very similar results to those in the
previous section. The CGs, when they are initially defined, have small
correlations which are indistinguishable in the figures from
those in the mother particle distribution, but at larger $k$ (of order
the inverse of the CG scale $\ell_{CG}$) they differ. We see also the
small differences at these scales between the Eulerian and Lagrangian
CG. As in the previous section we see that the relative differences
between the evolved particle distribution decreases in time, with a
greater convergence as non-linearity develops at smaller scales. Each
PS must converge to the corresponding Poisson noise level at large
$k$, which is simply the inverse of the particle density. Outside this
range the PS of each CG interpolates rapidly, in the evolved particle
distributions, to the PS of the mother distribution.
  
\begin{figure*}
  \includegraphics[width=\textwidth]{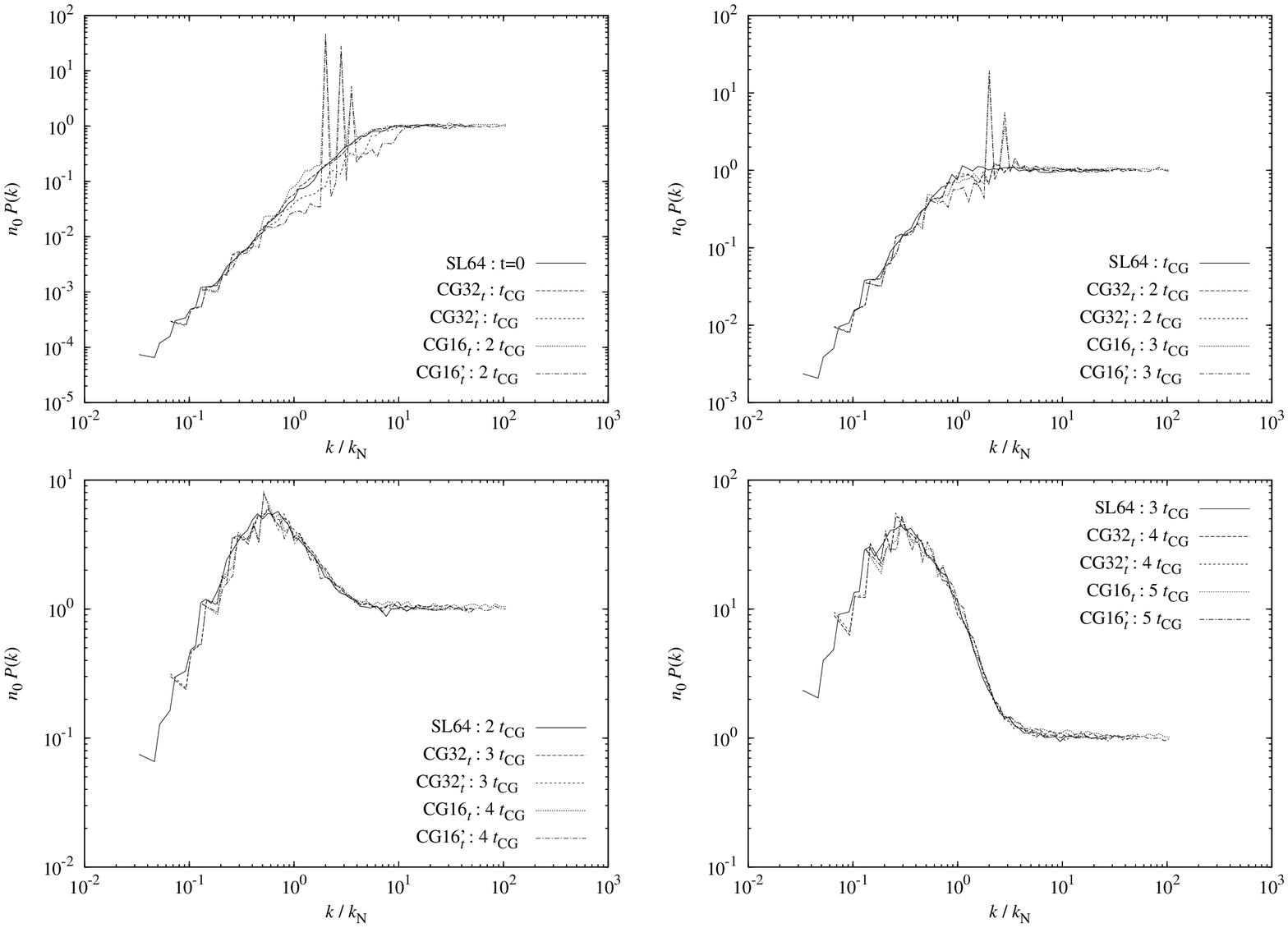}
  \caption{Comparison of the PS in SL64 at times $0,1,2,3$ in units of
    $t_\text{CG}$ with the PS in CG32$_t^\prime$ at times $1,2,3,4$  and the
    PS in CG16$_t^\prime$ at times $2,3,4,5$. We have plotted $n_0 P(k)$ as a 
function of $k/k_N$, where $k_N$ is the Nyquist frequency for the given
distribution. In the approximation that the ZA is valid at the scale 
of coarse-graining, all the plots should overlap.
\label{fig:CGintimePSRescaled}
 }
\end{figure*}

\begin{figure*}
\psfrag{roverell}{$r/\ell$}
  \includegraphics[width=\textwidth]{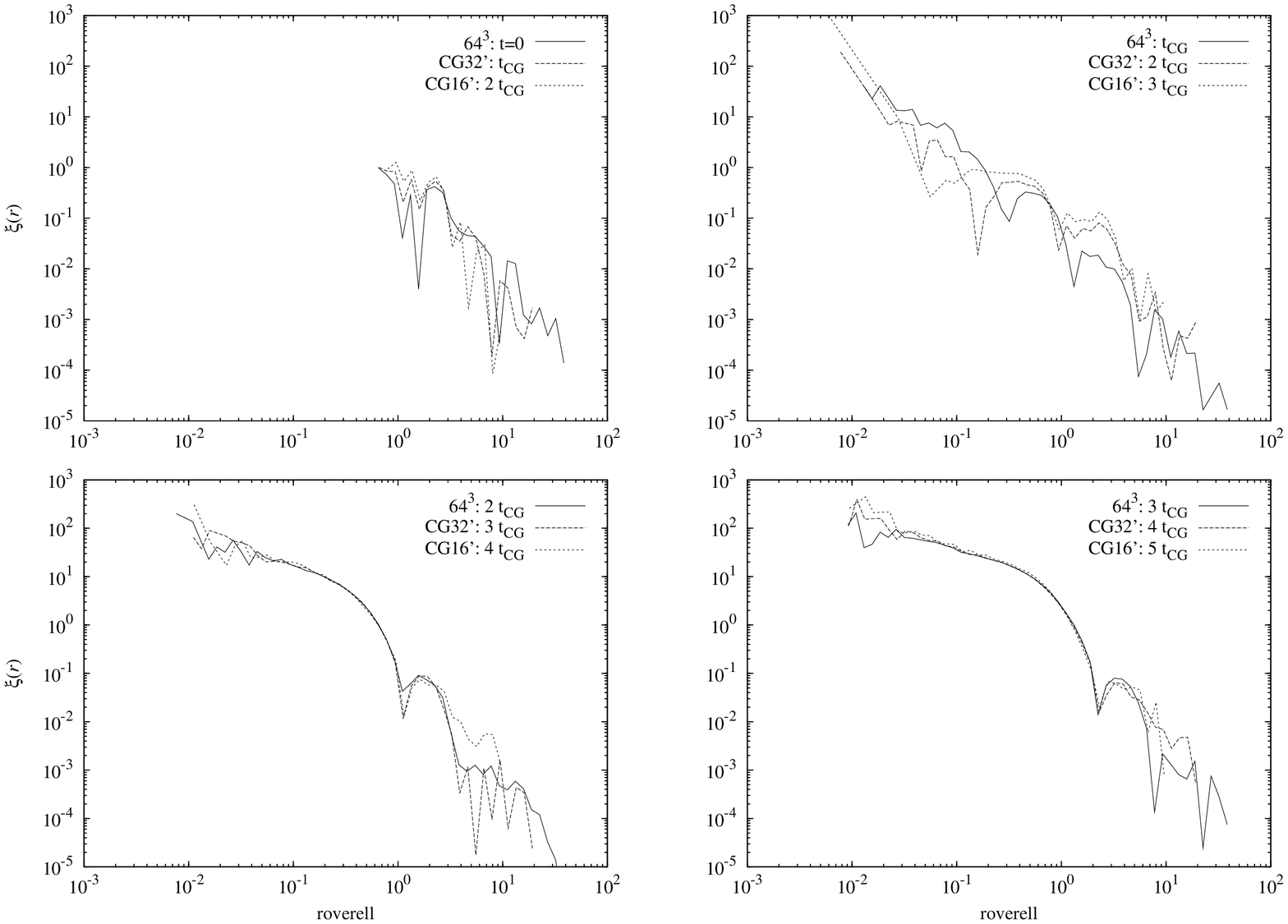}
  \caption{Comparison of $\xi(r)$ in SL64 at times $0,1,2,3$ in units of
    $t_\text{CG}$ with $\xi(r)$ in CG32$'_t$ at times $1,2,3,4$  and 
    in CG16$'_t$ at times $2,3,4,5$. As in the previous figure, the
approriate rescaling of lengths on the $x$-axis has been done
so that these functions should overlap in the approximation that
ZA is exact at (and above) the CG scale.
\label{fig:CGintimeXiRescaled}
}
\end{figure*}

As discussed above, the CG at a finite time we have defined 
is useful for understanding the self-similar properties of the 
asymptotic evolution from SL initial conditions\cite{sl1}. 
Indeed it has the property that, if the ZA is exact, the 
CG particle distribution defined in this way is identical (statistically)
to the initial condition up to a rescaling of length scales.
The ``self-similarity'' is then a direct consequence of the fact 
that the evolution before the onset of non-linearity on a 
given scale depends to a very good approximation only on the 
initial fluctuations at these scales, which themselves have (by
construction) no intrinsic length scale, and the purely 
fluid evolution of these fluctuations described by the ZA.

In this respect it is interesting to check directly to what 
extent the CG mass distribution actually is a rescaling 
of the mother distribution,
i.e. to what extent the ZA is an accurate approximation to the 
motion of the center of masses at the scale $\ell_{CG}$,
by direct comparison of the particle distributions. Further 
it is interesting to see how well the relation 
Eq.~(\ref{self-sim}) is obeyed. In Figs.~\ref{fig:CGintimePSRescaled}
and ~\ref{fig:CGintimeXiRescaled} we show these comparisons.
In the first the dimensionless combination $n_0P(k)$ is
plotted as a function of the $k$ wavenumber, rescaled as
required to the Nyquist frequency in each particle distribution
(this corresponds to a rescaling of length scales by
$\alpha$). In Fig.~\ref{fig:CGintimeXiRescaled} likewise
the $x$ axis has been rescaled in the same manner
\footnote{See the footnote in discussion of Fig.~\ref{fig:CGt0ps} 
above for the explanation of why not all the predicted Bragg peaks 
are visible in the plot.}. We see that the particle distributions are indeed 
initially  very similar --- except for some visible differences
close to the Nyquist frequency, due to deviations
from the ZA at these small scales ---  and in time
become more similar as the evolution develops.

\section{Conclusions and discussion} 
\label{Discussion and conclusions} 

We have studied the evolution of an infinite SL of points under its
self-gravity by comparing it with that of its particle CGs, using two
different prescriptions for the latter. The primary aim is to get
insight into the dynamics of the non-linear gravitational clustering
in this system, which has been shown in \cite{sl1} to be very similar
qualitatively to that in simulations of structure formation in
cosmology.  We first summarize the main conclusions we draw from our
study. We then discuss one specific question, which is important with
respect to the theoretical interpretation of these results: the
definition of a limit in which a continuum theoretical description
should be appropriate to fully understand the evolution (in the linear
{\it and} non-linear regime) of this system. This question is of
particular relevance in the context of qualitatively similar
simulations in cosmology, as such simulations are useful insofar as
they reproduce the evolution of such a limit. How they do so is,
however,currently poorly understood. Finally we briefly mention some
directions for further work.

\subsection{Summary of results}
Our principal results and conclusions are the following:

\begin{itemize}

\item  In the {\it numerical simulations} the coarse-grained 
particle distributions are observed to evolve to give, after a sufficient 
time, two-point correlation properties which agree well, over the 
range of scales simulated, with those in the original distribution. 
Indeed both the original system and its coarse-grainings 
converge toward a simple dynamical scaling (``self-similar'') 
behavior {\it with the same amplitude}. The characteristic time 
required for the CG system to begin to reproduce the clustering in the 
original particle distribution at scales {\it below} the CG scale 
increases as the latter scale does.  As discussed 
in Sec.~\ref{Results of numerical simulations} these observations are all 
very much in line qualitatively, and to a first approximation (see below)
quantitatively, with the qualitative picture of the evolution of clustering 
widely accepted in cosmology: the CG distributions share the same 
fluctuations at large scales and it is these initial fluctuations alone, 
to a very good approximation, which determine the correlations which
develop at smaller scales at later times.
 
\item  Once particles begin to fall on one another in any
of the SL simulations, there is a phase in which 
very significant non-linear correlations develop due to 
interactions between NN pairs of particles. Further we have 
seen, as in \cite{Baertschiger:2004tx, sl1}, that the {\it form} of 
the two-point correlation function which develops
in this phase is very similar to that observed, in the same range
of amplitude, in the asymptotic scaling regime at later times.
Thus it appears that it is always possible to choose a CG of the 
original system, which reproduces quite well the non-linear
correlations in the original system with this ``early time'',
explicitly discrete, dynamics of ``macro-particles'' of the
CG distribution.

This provides a simple physical picture/dynamical model for the 
generation of the non-linear correlation function in the relevant
range (up to an amplitude of $\xi$ in the range of
$10-10^2$). We will describe more quantitatively in a 
forthcoming paper \cite{sl3} the exact form of the 
non-linear correlation function given by this model.

This finding is {\it prima facie} very different, in spirit 
and substance, to any existing explanations of the dynamics
giving rise to non-linear correlations in N body simulations
in cosmology. In this context theoretical modeling invariably
assumes that the non-linear correlations observed in simulations
in this range should be understood in the framework of a 
continuum Vlasov limit, in which a mean-field approximation of the
gravitational field is appropriate. Indeed the fact that self-similarity
is observed, with a behavior independent of the particle density,
is usually taken as an indication that such a continuum description
is appropriate. Our model is manifestly not of this type (while
also consistent with the amplitudes of the correlation function
being independent of particle density).
There is, of course, not necessarily a contradiction: it may be that 
our simple discrete model is simply a 
sufficiently good approximation to the continuum model\footnote{
In \cite{Baertschiger:2002tk} we have presented an analysis of
the force distribution in cosmological simulations at the time
of formation of the first non-linear correlations, showing that
NN interactions also dominate in this case. Thus we do not
expect that either the different initial conditions, or the
fact that there is expansion of the spatial background, should
qualitatively change the early time dynamics with respect to that
of the system discussed here.}. We will return to a 
discussion of this point below.

\item A closer examination of the results shows that there are small,
but observable, differences in amplitudes between the original SL and
it CGs. These may be explained using a more accurate description of
the early time evolution of the system, given by the perturbative PLT
treatment of \cite{joyce_05, marcos_06}, valid up to the onset
of the NN interaction phase. In this framework the fluid limit
criterion Eq.~(\ref{fluid-collapse-CG}) for the end of this phase may
be replaced by a more accurate criterion, taking into account the
dependence of the rate of evolution of modes of the displacement field
at small scales on particle density (and their wavelength). Generically
the effect of discreteness (i.e. the difference with respect to the
fluid limit) is to slow down the growth of modes at smaller scales
(comparable to the inverse of the inter-particle spacing). Thus the
time of exit from the PLT phase is longer than given by
Eq.~(\ref{fluid-collapse-CG}).  In our numerical simulations a
resultant ``lagging'' of the non-linear correlation function in the CG
simulation is just discernible at early times. These effects are
small, and visible only at very early times, because the value of
$\delta$ used, albeit small compared to unity, is not so small. In the
limit that the normalized shuffling $\delta$ becomes arbitrarily
small, however, these differences between the discrete and fluid
system can become arbitrarily large.

\item Since the coarse-grainings defined on the initial conditions
give $\delta \propto (\ell/ \ell_{CG})^{5/2}$, differences between
the full PLT evolution and the fluid evolution will, for the
reasons explained in the previous paragraph, diverge in 
the limit $(\ell/ \ell_{CG}) \rightarrow 0$. The coarse-grainings 
defined at finite times avoids this feature by construction: it ensures 
that the CG system has approximately the same $\delta$, so that the time
spent in the phase in which discreteness effects accumulate
as described by PLT is identical to that in the mother 
particle distribution. Thus the use of a CG distribution, evolving
in its ``early'' phase, to reproduce the non-linear 
correlations in the original distribution would be 
expected to work in this case even for arbitrarily large
times (which require arbitrarily large values of $\ell_{CG}$).

\item The origin of self-similarity observed in the evolution of the
system, in which the temporal evolution of the correlation function is
equivalent to a rescaling of spatial variables, is nicely illustrated
using the CG at finite times. Indeed insofar as the system, for its
further evolution, can be approximated by its CG defined in this way,
one obtains self-similarity. We have seen that there are thus two
ingredients giving this behavior. Firstly the non-linear correlations
which develop at a given scale depend on the initial conditions
only through the fluctuations present initially at larger scales (i.e. the
same as the first point above). Secondly these larger scales are well described
at early times by the Zeldovich approximation, i.e., by the purely
fluid limit in which the evolution is independent of the particle
density.

\end{itemize}

\subsection{The continuum limit}

As noted in the introduction, one of the motivations for studying
coarse-grainings like those we have considered here is that such 
a study may help to clarify what properties of the evolved SL may be 
understood as valid in a continuum limit: the idea is that 
such properties should be invariant under coarse-graining, i.e., 
invariant with respect to changes in the small scale discrete 
structure  of the systems. The fact that we have found that 
the evolution observed in the original system is not completely
invariant in this sense, and indeed that the different coarse-graining 
procedures give different results, shows that this evolution is 
not fully representative of such a continuum limit.  The question of 
such a limit is of relevance if one wants to  determine whether, in 
particular, the {\it form} of the correlation function in the 
self-similar should be derivable with such a continuum theory.
As mentioned above, this is particularly important to understand
in the context of simulations of this type in cosmology, as 
the goal of the N body method is to reproduce such a limit.
 
To address this question we need to make more precise what we mean 
by ``continuum limit''. There is in fact, of course, no unique way
of defining such a limit. Implicitly above we have been supposing
a limit in which the inter-particle spacing $\ell$ goes to zero
in some appropriate manner, so that a hydrodynamic type description
of the system becomes valid. More specifically, in cosmology and 
more generally for systems with long range interactions, one is 
interested in identifying the Vlasov limit, in which the system is 
described by a one particle phase space density obeying
a Liouville equation in which the force is calculated in 
a self-consistent mean field approximation (see, e.g. 
\cite{peebles, saslaw2000}).

The SL system we have studied is characterized \cite{sl1} by a single 
{\it dimensionless} parameter: the normalized shuffling parameter 
$\delta$. Indeed the evolution we have studied is, as discussed 
in \cite{sl1}, independent of both the  system size and force softening. 
Two SL with the same $\delta$, but different $\ell$, are then 
equivalent because of the scale free nature of gravity. To give meaning 
to a limit $\ell \rightarrow 0$ we must, therefore, evidently introduce 
at least one additional length scale. We note that in the discussion of the ZA
above, in Sec.~\ref{The Zeldovich Approximation}, we have derived
it as a continuum limit by fixing the wave length of the mode 
considered and taking $\ell \rightarrow 0$. In this way we 
established that long wavelength modes of the displacement field
behave as derived from a set of fluid equations for the 
self-gravitating fluid. The SL system {\it as a whole}, however,
cannot have any continuum limit in the desired sense unless
we introduce an additional length scale in the system.

In \cite{braun+hepp} it is shown formally that such a Vlasov
description is recovered in the limit that $N \rightarrow \infty$ for 
a class of  long-range interacting system of $N$ particles, provided 
their interaction potentials are regulated at small separations. 
This suggests that one way in which we should recover such a 
continuum description is by introducing such a regularization, 
characterized by a scale $\varepsilon$. Then one can take 
$\ell \rightarrow 0$  at fixed $\varepsilon$ 
(and then, in principle $\varepsilon \rightarrow 0$).  

Following what was said above, we expect that in this limit
it should be possible to define coarse-graining procedures
like we have done which leave the evolution of the system
invariant. It is not difficult to see that this should indeed
be the case. Let us assume that the force falls rapidly to zero 
at separations less than $\varepsilon$. The PLT treatment 
can be applied and will give a spectrum of modes which
are modified with respect to full gravity for 
$k\varepsilon > 1$. In the limit $\ell/\varepsilon \rightarrow 0$ 
these modifications will become independent of $\ell$.
Thus the evolution of the displacement field will be
independent of $\ell$. As the CG defined are just a 
modification of this latter scale, they will have no
effect on the evolution (provided
$\ell_{CG}/\varepsilon \ll 1$). In practice this means
that we can coarse-grain the system of finite $\ell$
(and $\ell \ll \varepsilon$) up to any $\ell_{CG}$,
provided $\ell_{CG} \ll \varepsilon$. 

An alternative way of defining the limit $\ell \rightarrow 0$ 
is by introducing a characteristic scale not in the force, but 
in the fluctuations in the system. The SL has no such scale, but 
it is easy to introduce one by defining instead the initial
displacement field so that it has such a cut-off: we 
derive it as the sum of modes in reciprocal space, with
all modes $k > k_c$ set equal to zero. We can then define
a continuum limit for the system as 
$k_c \ell \rightarrow 0$. It is not difficult to see that
a similar reasoning to that used for the previous case
leads to the conclusion that coarse-graining like those
we have defined now change negligibly the evolution of
the system provided $k_c \ell_{CG} \ll 1$. 

We note that there is one intrinsic problem with the latter
prescription: it works because the scale $k_c$ 
is the one characterizing the shortest wavelength fluctuation
in the system. This is only true, however, at the initial
time, as gravity will progressively (and very efficiently)
create fluctuations at smaller scales. Thus the continuum
limit will only be valid for some finite time. For the
former case this will not be the case as the finite value
of $\varepsilon$ will place a time-independent lower bound 
on the scales at which fluctuations develop. There is thus
an intrinsic time independent separation between the 
scales of inhomogeneity and the particle scale. 

This discussion tells us then how we should extrapolate our
simulations toward a well defined continuum (Vlasov) theory 
to determine which results are valid in this limit: we 
should increase particle density until $\ell \ll \varepsilon$. 
Manifestly this will ``kill'' the discrete particle dynamics
which we have used in our model to explain the {\it form} of
the non-linear two-point correlation function.  If this form
is indeed that of the continuum limit it should be observed
in simulations with $\ell \ll \varepsilon$ when non-linear
correlations develop for $r \gg \varepsilon$. To perform 
simulations to test whether this is so requires that
there be a reasonable range of scale 
$r \gg \epsilon \gg \ell$ well inside the simulation
box, which implies a particle number much greater than 
that in the largest $64^3$ simulation discussed here
(Taking, e.g., $\ell=0.2 \varepsilon$, $\varepsilon$ is 
already almost one tenth of the box size). 
With the particle numbers, in the range $10^8-10^{10}$,
now being accessed by several gravitational N body 
simulation groups (see, e.g., \cite{springel_05, maccioetal_06, jingetal_06})
such numerical tests are, however, in principle feasible and would 
yield clear conclusions on this very basic question.

\acknowledgments{
We thank the ``Centro Ricerche e Studi E. Fermi'' (Roma) for the use
of a super-computer for numerical calculations, the
MIUR-PRIN05 project on ``Dynamics and thermodynamics of systems with
long range interactions" for financial support.  M.\,J. thanks the
Istituto dei Sistemi Complessi for its kind hospitality during May
2006 and October 2006. We acknowledge Bruno Marcos for interesting
discussions and comments.}


\begin{thebibliography}{36}
\expandafter\ifx\csname natexlab\endcsname\relax\def\natexlab#1{#1}\fi
\expandafter\ifx\csname bibnamefont\endcsname\relax
  \def\bibnamefont#1{#1}\fi
\expandafter\ifx\csname bibfnamefont\endcsname\relax
  \def\bibfnamefont#1{#1}\fi
\expandafter\ifx\csname citenamefont\endcsname\relax
  \def\citenamefont#1{#1}\fi
\expandafter\ifx\csname url\endcsname\relax
  \def\url#1{\texttt{#1}}\fi
\expandafter\ifx\csname urlprefix\endcsname\relax\def\urlprefix{URL }\fi
\providecommand{\bibinfo}[2]{#2}
\providecommand{\eprint}[2][]{\url{#2}}

\bibitem[{\citenamefont{Baertschiger et~al.}(2006)\citenamefont{Baertschiger,
  Joyce, Gabrielli, and Sylos~Labini}}]{sl1}
\bibinfo{author}{\bibfnamefont{T.}~\bibnamefont{Baertschiger}},
  \bibinfo{author}{\bibfnamefont{M.}~\bibnamefont{Joyce}},
  \bibinfo{author}{\bibfnamefont{A.}~\bibnamefont{Gabrielli}},
  \bibnamefont{and}
  \bibinfo{author}{\bibfnamefont{F.}~\bibnamefont{Sylos~Labini}},
  \bibinfo{journal}{Phys. Rev.} \textbf{\bibinfo{volume}{E.}}
  (\bibinfo{year}{2006}), \bibinfo{note}{to appear}, \eprint{cond-mat/0607396}.

\bibitem[{\citenamefont{Cardy}(1996)}]{cardy}
\bibinfo{author}{\bibfnamefont{J.}~\bibnamefont{Cardy}},
  \emph{\bibinfo{title}{Scaling and Renormalization in Statistical Physics}},
\bibinfo{address}{Cambidge Lecture Notes in Physics},  
(\bibinfo{publisher}{Cambridge}, \bibinfo{year}{1996}).

\bibitem[{\citenamefont{Gaite and Dominguez}(2006)}]{gaite_dominguez_scaling}
\bibinfo{author}{\bibfnamefont{J.}~\bibnamefont{Gaite}} \bibnamefont{and}
  \bibinfo{author}{\bibfnamefont{A.}~\bibnamefont{Dominguez}}
  (\bibinfo{year}{2006}), \eprint{cond-mat/0610886}.

\bibitem[{\citenamefont{Gaite}(2001)}]{gaite_proc2000_renormalization}
\bibinfo{author}{\bibfnamefont{J.}~\bibnamefont{Gaite}}, \bibinfo{journal}{Int.
  J. Mod. Phys.} \textbf{\bibinfo{volume}{A16}}, \bibinfo{pages}{2041}
  (\bibinfo{year}{2001}), \eprint{cond-mat/0101219}.

\bibitem[{\citenamefont{Sota et~al.}(1998)\citenamefont{Sota, Kobayashi, Maeda,
  Kurokawa, Morikawa, and Nakamichi}}]{Sota_etal_renormalization}
\bibinfo{author}{\bibfnamefont{Y.}~\bibnamefont{Sota}},
  \bibinfo{author}{\bibfnamefont{T.}~\bibnamefont{Kobayashi}},
  \bibinfo{author}{\bibfnamefont{K.}~\bibnamefont{Maeda}},
  \bibinfo{author}{\bibfnamefont{T.}~\bibnamefont{Kurokawa}},
  \bibinfo{author}{\bibfnamefont{M.}~\bibnamefont{Morikawa}}, \bibnamefont{and}
  \bibinfo{author}{\bibfnamefont{A.}~\bibnamefont{Nakamichi}},
  \bibinfo{journal}{Phys. Rev.} \textbf{\bibinfo{volume}{D58}},
  \bibinfo{pages}{043502} (\bibinfo{year}{1998}), \eprint{gr-qc/9801083}.

\bibitem[{\citenamefont{Semelin et~al.}(1999)\citenamefont{Semelin, de~Vega,
  Sanchez, and Combes}}]{Semelin_renormalization}
\bibinfo{author}{\bibfnamefont{B.}~\bibnamefont{Semelin}},
  \bibinfo{author}{\bibfnamefont{H.}~\bibnamefont{de~Vega}},
  \bibinfo{author}{\bibfnamefont{N.}~\bibnamefont{Sanchez}}, \bibnamefont{and}
  \bibinfo{author}{\bibfnamefont{F.}~\bibnamefont{Combes}},
  \bibinfo{journal}{Phys. Rev.} \textbf{\bibinfo{volume}{D59}},
  \bibinfo{pages}{125021} (\bibinfo{year}{1999}), \eprint{astro-ph/9812467}.

\bibitem[{\citenamefont{Antonov}(2004)}]{Antonov_scaling}
\bibinfo{author}{\bibfnamefont{N.}~\bibnamefont{Antonov}},
  \bibinfo{journal}{Phys. Rev. Lett.} \textbf{\bibinfo{volume}{92}},
  \bibinfo{pages}{161101} (\bibinfo{year}{2004}), \eprint{astro-ph/0308369}.

\bibitem[{\citenamefont{Peebles}(1985)}]{Peebles_renormalization}
\bibinfo{author}{\bibfnamefont{P.}~\bibnamefont{Peebles}},
  \bibinfo{journal}{Astrophys. J.} \textbf{\bibinfo{volume}{297}},
  \bibinfo{pages}{350} (\bibinfo{year}{1985}).

\bibitem[{\citenamefont{Peebles and Couchman}(1998)}]{Peebles+couchman_1995}
\bibinfo{author}{\bibfnamefont{P.}~\bibnamefont{Peebles}} \bibnamefont{and}
  \bibinfo{author}{\bibfnamefont{H.}~\bibnamefont{Couchman}},
  \bibinfo{journal}{Astrophys. J.} \textbf{\bibinfo{volume}{497}},
  \bibinfo{pages}{499} (\bibinfo{year}{1998}), \eprint{astro-ph/9708230}.

\bibitem[{\citenamefont{Splinter et~al.}(1998)\citenamefont{Splinter, Melott,
  Shandarin, and Suto}}]{splinter}
\bibinfo{author}{\bibfnamefont{R.~J.} \bibnamefont{Splinter}},
  \bibinfo{author}{\bibfnamefont{A.~L.} \bibnamefont{Melott}},
  \bibinfo{author}{\bibfnamefont{S.~F.} \bibnamefont{Shandarin}},
  \bibnamefont{and} \bibinfo{author}{\bibfnamefont{Y.}~\bibnamefont{Suto}},
  \bibinfo{journal}{Astrophys. J.} \textbf{\bibinfo{volume}{497}},
  \bibinfo{pages}{38} (\bibinfo{year}{1998}).

\bibitem[{\citenamefont{Marcos et~al.}(2006)\citenamefont{Marcos, Baertschiger,
  Joyce, Gabrielli, and Sylos~Labini}}]{marcos_06}
\bibinfo{author}{\bibfnamefont{B.}~\bibnamefont{Marcos}},
  \bibinfo{author}{\bibfnamefont{T.}~\bibnamefont{Baertschiger}},
  \bibinfo{author}{\bibfnamefont{M.}~\bibnamefont{Joyce}},
  \bibinfo{author}{\bibfnamefont{A.}~\bibnamefont{Gabrielli}},
  \bibnamefont{and}
  \bibinfo{author}{\bibfnamefont{F.}~\bibnamefont{Sylos~Labini}},
  \bibinfo{journal}{Phys. Rev} \textbf{\bibinfo{volume}{D73}},
  \bibinfo{pages}{103507} (\bibinfo{year}{2006}), \eprint{astro-ph/0601479}.

\bibitem[{\citenamefont{Joyce et~al.}(2005)\citenamefont{Joyce, Marcos,
  Gabrielli, Baertschiger, and Sylos~Labini}}]{joyce_05}
\bibinfo{author}{\bibfnamefont{M.}~\bibnamefont{Joyce}},
  \bibinfo{author}{\bibfnamefont{B.}~\bibnamefont{Marcos}},
  \bibinfo{author}{\bibfnamefont{A.}~\bibnamefont{Gabrielli}},
  \bibinfo{author}{\bibfnamefont{T.}~\bibnamefont{Baertschiger}},
  \bibnamefont{and}
  \bibinfo{author}{\bibfnamefont{F.}~\bibnamefont{Sylos~Labini}},
  \bibinfo{journal}{Phys. Rev. Lett.} \textbf{\bibinfo{volume}{95}},
  \bibinfo{pages}{011304} (\bibinfo{year}{2005}), \eprint{astro-ph/0504213}.

\bibitem[{\citenamefont{Gabrielli et~al.}(2006)\citenamefont{Gabrielli,
  Baertschiger, Joyce, Marcos, and Sylos~Labini}}]{gabrielli_06}
\bibinfo{author}{\bibfnamefont{A.}~\bibnamefont{Gabrielli}},
  \bibinfo{author}{\bibfnamefont{T.}~\bibnamefont{Baertschiger}},
  \bibinfo{author}{\bibfnamefont{M.}~\bibnamefont{Joyce}},
  \bibinfo{author}{\bibfnamefont{B.}~\bibnamefont{Marcos}}, \bibnamefont{and}
  \bibinfo{author}{\bibfnamefont{F.}~\bibnamefont{Sylos~Labini}},
  \bibinfo{journal}{Phys. Rev.} \textbf{\bibinfo{volume}{E74}},
  \bibinfo{pages}{021110} (\bibinfo{year}{2006}), \eprint{cond-mat/0603124}.

\bibitem[{\citenamefont{Gabrielli
  et~al.}(2004{\natexlab{a}})\citenamefont{Gabrielli, Sylos~Labini, Joyce, and
  Pietronero}}]{book}
\bibinfo{author}{\bibfnamefont{A.}~\bibnamefont{Gabrielli}},
  \bibinfo{author}{\bibfnamefont{F.}~\bibnamefont{Sylos~Labini}},
  \bibinfo{author}{\bibfnamefont{M.}~\bibnamefont{Joyce}}, \bibnamefont{and}
  \bibinfo{author}{\bibfnamefont{L.}~\bibnamefont{Pietronero}},
  \emph{\bibinfo{title}{Statistical Physics for Cosmic Structures}}
  (\bibinfo{publisher}{Springer}, \bibinfo{year}{2004}{\natexlab{a}}).

\bibitem[{\citenamefont{Baertschiger and
  Sylos~Labini}(2004)}]{Baertschiger:2004tx}
\bibinfo{author}{\bibfnamefont{T.}~\bibnamefont{Baertschiger}}
  \bibnamefont{and}
  \bibinfo{author}{\bibfnamefont{F.}~\bibnamefont{Sylos~Labini}},
  \bibinfo{journal}{Phys. Rev.} \textbf{\bibinfo{volume}{D69}},
  \bibinfo{pages}{123001} (\bibinfo{year}{2004}), \eprint{astro-ph/0401238}.

\bibitem[{\citenamefont{Gabrielli}(2004)}]{andrea}
\bibinfo{author}{\bibfnamefont{A.}~\bibnamefont{Gabrielli}},
  \bibinfo{journal}{Phys. Rev.} \textbf{\bibinfo{volume}{E70}},
  \bibinfo{pages}{066131} (\bibinfo{year}{2004}), \eprint{cond-mat/0409594}.

\bibitem[{\citenamefont{{Efstathiou} et~al.}(1988)\citenamefont{{Efstathiou},
  {Frenk}, {White}, and {Davis}}}]{efstathiou_88}
\bibinfo{author}{\bibfnamefont{G.}~\bibnamefont{{Efstathiou}}},
  \bibinfo{author}{\bibfnamefont{C.~S.} \bibnamefont{{Frenk}}},
  \bibinfo{author}{\bibfnamefont{S.~D.~M.} \bibnamefont{{White}}},
  \bibnamefont{and} \bibinfo{author}{\bibfnamefont{M.}~\bibnamefont{{Davis}}},
  \bibinfo{journal}{Mon. Not. R. Astron. Soc.} \textbf{\bibinfo{volume}{235}},
  \bibinfo{pages}{715} (\bibinfo{year}{1988}).

\bibitem[{\citenamefont{Peebles}(1980)}]{peebles}
\bibinfo{author}{\bibfnamefont{P.~J.~E.} \bibnamefont{Peebles}},
  \emph{\bibinfo{title}{{The Large-Scale Structure of the Universe}}}
  (\bibinfo{publisher}{Princeton University Press}, \bibinfo{year}{1980}).

\bibitem[{\citenamefont{Pines}(1963)}]{pines}
\bibinfo{author}{\bibfnamefont{D.}~\bibnamefont{Pines}},
  \emph{\bibinfo{title}{Elementary Excitations in Solids}}
  (\bibinfo{publisher}{Benjamin, New York}, \bibinfo{year}{1963}).

\bibitem[{\citenamefont{Zeldovich}(1970)}]{zeldovich_70}
\bibinfo{author}{\bibfnamefont{Y.~B.} \bibnamefont{Zeldovich}},
  \bibinfo{journal}{Astron. Astrophys.} \textbf{\bibinfo{volume}{5}},
  \bibinfo{pages}{84} (\bibinfo{year}{1970}).

\bibitem[{\citenamefont{Buchert}(1992)}]{buchert2}
\bibinfo{author}{\bibfnamefont{T.}~\bibnamefont{Buchert}},
  \bibinfo{journal}{Mon. Not. R. Astron. Soc.} \textbf{\bibinfo{volume}{254}},
  \bibinfo{pages}{729} (\bibinfo{year}{1992}).

\bibitem[{\citenamefont{Zeldovich}(1965)}]{zeldovich-k4}
\bibinfo{author}{\bibfnamefont{Y.~B.} \bibnamefont{Zeldovich}},
  \bibinfo{journal}{Adv. Astron.} \textbf{\bibinfo{volume}{3}},
  \bibinfo{pages}{241} (\bibinfo{year}{1965}).

\bibitem[{\citenamefont{Zeldovich and Novikov}(1983)}]{zeldo}
\bibinfo{author}{\bibfnamefont{Y.}~\bibnamefont{Zeldovich}} \bibnamefont{and}
  \bibinfo{author}{\bibfnamefont{I.}~\bibnamefont{Novikov}},
  \emph{\bibinfo{title}{Relativistic Astrophysics}} (\bibinfo{publisher}{Univ.
  Chicago Press}, \bibinfo{address}{Chicago}, \bibinfo{year}{1983}).

\bibitem[{\citenamefont{Gabrielli
  et~al.}(2004{\natexlab{b}})\citenamefont{Gabrielli, Joyce, Marcos, and
  Viot}}]{gabrielli_etal_04}
\bibinfo{author}{\bibfnamefont{A.}~\bibnamefont{Gabrielli}},
  \bibinfo{author}{\bibfnamefont{M.}~\bibnamefont{Joyce}},
  \bibinfo{author}{\bibfnamefont{B.}~\bibnamefont{Marcos}}, \bibnamefont{and}
  \bibinfo{author}{\bibfnamefont{P.}~\bibnamefont{Viot}},
  \bibinfo{journal}{Europhys. Lett.} \textbf{\bibinfo{volume}{66}},
  \bibinfo{pages}{1} (\bibinfo{year}{2004}{\natexlab{b}}),
  \eprint{astro-ph/0303169}.

\bibitem[{{\textnormal{\texttt{www.mpa-garching.mpg.de/galform/gadget/index.sh%
tml}}}()}]{gadget}
{\textnormal{\texttt{www.mpa-garching.mpg.de/galform/gadget/index.shtml}}}.

\bibitem[{\citenamefont{Joyce and Marcos}(2004)}]{michael_bruno}
\bibinfo{author}{\bibfnamefont{M.}~\bibnamefont{Joyce}} \bibnamefont{and}
  \bibinfo{author}{\bibfnamefont{B.}~\bibnamefont{Marcos}}
  (\bibinfo{year}{2004}), \eprint{astro-ph/0410451}.

\bibitem[{\citenamefont{Padmanabhan}(2002)}]{padmanabhan_short}
\bibinfo{author}{\bibfnamefont{T.}~\bibnamefont{Padmanabhan}},
  \bibinfo{journal}{Khagol} \textbf{\bibinfo{volume}{51}}, \bibinfo{pages}{5}
  (\bibinfo{year}{2002}), \eprint{astro-ph/0308499}.

\bibitem[{\citenamefont{Little et~al.}(1991)\citenamefont{Little, Weinberg, and
  Park}}]{Little+weinberg+park_1991}
\bibinfo{author}{\bibfnamefont{B.}~\bibnamefont{Little}},
  \bibinfo{author}{\bibfnamefont{D.}~\bibnamefont{Weinberg}}, \bibnamefont{and}
  \bibinfo{author}{\bibfnamefont{C.}~\bibnamefont{Park}},
  \bibinfo{journal}{Mon. Not. R. Astron. Soc.} \textbf{\bibinfo{volume}{253}},
  \bibinfo{pages}{295} (\bibinfo{year}{1991}).

\bibitem[{\citenamefont{Peacock and Dodds}(1996)}]{peacock}
\bibinfo{author}{\bibfnamefont{J.~A.} \bibnamefont{Peacock}} \bibnamefont{and}
  \bibinfo{author}{\bibfnamefont{S.~J.} \bibnamefont{Dodds}},
  \bibinfo{journal}{Mon. Not. R. Astron. Soc.} \textbf{\bibinfo{volume}{280}},
  \bibinfo{pages}{L19} (\bibinfo{year}{1996}).

\bibitem[{\citenamefont{Baertschiger et~al.}(2007)\citenamefont{Baertschiger,
  Joyce, Gabrielli, and Sylos~Labini}}]{sl3}
\bibinfo{author}{\bibfnamefont{T.}~\bibnamefont{Baertschiger}},
  \bibinfo{author}{\bibfnamefont{M.}~\bibnamefont{Joyce}},
  \bibinfo{author}{\bibfnamefont{A.}~\bibnamefont{Gabrielli}},
  \bibnamefont{and}
  \bibinfo{author}{\bibfnamefont{F.}~\bibnamefont{Sylos~Labini}}
  (\bibinfo{year}{2007}), \bibinfo{note}{in preparation}.

\bibitem[{\citenamefont{Baertschiger et~al.}(2002)\citenamefont{Baertschiger,
  Joyce, and Sylos~Labini}}]{Baertschiger:2002tk}
\bibinfo{author}{\bibfnamefont{T.}~\bibnamefont{Baertschiger}},
  \bibinfo{author}{\bibfnamefont{M.}~\bibnamefont{Joyce}}, \bibnamefont{and}
  \bibinfo{author}{\bibfnamefont{F.}~\bibnamefont{Sylos~Labini}},
  \bibinfo{journal}{Astrophys. J.} \textbf{\bibinfo{volume}{581}},
  \bibinfo{pages}{L63} (\bibinfo{year}{2002}), \eprint{astro-ph/0203087}.

\bibitem[{\citenamefont{Saslaw}(2000)}]{saslaw2000}
\bibinfo{author}{\bibfnamefont{W.}~\bibnamefont{Saslaw}},
  \emph{\bibinfo{title}{{The Distribution of the Galaxies}}}
  (\bibinfo{publisher}{Cambridge University Press}, \bibinfo{year}{2000}).

\bibitem[{\citenamefont{Braun and Hepp}(1977)}]{braun+hepp}
\bibinfo{author}{\bibfnamefont{W.}~\bibnamefont{Braun}} \bibnamefont{and}
  \bibinfo{author}{\bibfnamefont{K.}~\bibnamefont{Hepp}},
  \bibinfo{journal}{Comm. Math. Phys.} \textbf{\bibinfo{volume}{56}},
  \bibinfo{pages}{101} (\bibinfo{year}{1977}).

\bibitem[{\citenamefont{Springel et~al.}(2005)}]{springel_05}
\bibinfo{author}{\bibfnamefont{V.}~\bibnamefont{Springel}}
  \bibnamefont{et~al.}, \bibinfo{journal}{Nature}
  \textbf{\bibinfo{volume}{435}}, \bibinfo{pages}{629} (\bibinfo{year}{2005}),
  \eprint{astro-ph/0504097}.

\bibitem[{\citenamefont{Jing et~al.}(2007)\citenamefont{Jing, Suto, and
  Mo}}]{jingetal_06}
\bibinfo{author}{\bibfnamefont{Y.}~\bibnamefont{Jing}},
  \bibinfo{author}{\bibfnamefont{Y.}~\bibnamefont{Suto}}, \bibnamefont{and}
  \bibinfo{author}{\bibfnamefont{H.}~\bibnamefont{Mo}},
  \bibinfo{journal}{Astrophys. J.}  (\bibinfo{year}{2007}), \bibinfo{note}{to
  appear}, \eprint{astro-ph/0610099}.

\bibitem[{\citenamefont{Maccio' et~al.}(2006)\citenamefont{Maccio', Dutton,
  van~den Bosch, Moore, Potter, Stadel, and Diemand}}]{maccioetal_06}
\bibinfo{author}{\bibfnamefont{A.~V.} \bibnamefont{Maccio'}},
  \bibinfo{author}{\bibfnamefont{A.~A.} \bibnamefont{Dutton}},
  \bibinfo{author}{\bibfnamefont{F.~C.} \bibnamefont{van~den Bosch}},
  \bibinfo{author}{\bibfnamefont{B.}~\bibnamefont{Moore}},
  \bibinfo{author}{\bibfnamefont{D.}~\bibnamefont{Potter}},
  \bibinfo{author}{\bibfnamefont{J.}~\bibnamefont{Stadel}}, \bibnamefont{and}
  \bibinfo{author}{\bibfnamefont{J.}~\bibnamefont{Diemand}}
  (\bibinfo{year}{2006}), \eprint{astro-ph/0608157}.

\end{thebibliography}


\end{document}